%% file: DI.tex
\documentclass[10pt, conference]{IEEEtran}
\usepackage{amsmath,amsfonts,amssymb,epsfig,epstopdf,url,array}

\usepackage{amsthm}
\usepackage{color,soul}
\usepackage{algorithmic}
\usepackage{balance}
\usepackage[ruled]{algorithm2e}
\usepackage{stmaryrd}
\usepackage[table]{xcolor}
\usepackage{centernot}
\usepackage{tikz}
\usepackage{booktabs}
\usepackage{pdfpages}
\usepackage{graphicx}
\usepackage{subfigure}
\usepackage[font=small,skip=8pt]{caption}
\usepackage{tabularx}
\usepackage{comment}
\usepackage{multirow}
\usepackage{multicol}
\usepackage{bigstrut}
\usepackage{pdflscape}
\usepackage{rotating}
\usepackage{float}
\usepackage{lipsum}

\usepackage{enumitem}
\usetikzlibrary{plotmarks,shapes,arrows,chains,hobby,backgrounds,calc,trees}
\usepackage{makecell}
\usepackage{dblfloatfix}    
\usepackage{pdfcomment} 
\usepackage{lipsum}
\usepackage{bm}

\definecolor{lightgray}{gray}{0.9}
\theoremstyle{definition}
\newtheorem{exmp}{Example}
\theoremstyle{definition}

\theoremstyle{definition}
\newtheorem{defn}{Definition}

\begin{document}

\title{Dependency-Aware Software Release Planning through Mining User Preferences}


\author{\IEEEauthorblockN{Davoud Mougouei, David M. W. Powers}
\IEEEauthorblockA{School of Computer Science, Engineering, and Mathematics
Flinders University\\
Adelaide, Australia\\
\{davoud.mougouei,david.powers\}@flinders.edu.au}
}

\maketitle
\clubpenalty = 10000
\widowpenalty = 10000
\displaywidowpenalty = 10000

\begin{abstract}
Considering user preferences is a determining factor in optimizing the value of a software release. This is due to the fact that user preferences for software features specify the values of those features and consequently determine the value of the release. Certain features of a software however, may encourage or discourage users to prefer (select or use) other features. As such, value of a software feature could be positively or negatively influenced by other features. Such influences are known as \textit{Value-related Feature (Requirement) Dependencies}. Value-related dependencies need to be considered in software release planning as they influence the value of the  optimal subset of the features selected by the release planning models. Hence, we have proposed considering value-related feature dependencies in software release planning through mining user preferences for software features. We have demonstrated the validity and practicality of the proposed approach by studying a real world software project.
\end{abstract}
\vspace{1em}


\input{introduction}
\input{related}
\input{mining}
\input{mining_gathering}

\input{mining_identification}
\input{mining_resampling}
\input{modeling}

\input{factoring}
\input{case}
\input{conclusion}

\bibliographystyle{abbrv}
\bibliography{ref}  
\end{document}

%% file: introduction.tex
\section{introduction}
\label{sec_introduction}

Software release planning aims to find an optimal subset of software features (requirements) with the highest value while keeping the cost within the the budget~\cite{bagnall_next_2001}. The term ``value" however, can be misleading as it may be interpreted as the estimated value of a software feature expressed in currency units ignoring the impact of user preferences on the value of that feature. In reality however, the value of a software feature is determined by the user preferences for that feature~\cite{racheva2010business}. This is for the reason that a software feature preferred by the majority of the users is more likely to return its estimated value. Hence, the term ``customer value" or ``user value" has been preferred by some of the existing works~\cite{racheva2010business} to emphasize the impact of the user preferences in determining the values of software features. 

User preferences for software features however may be motivated by other features of the software. As such, the value of a software feature can be influenced by user preferences for other features. Such influences have been described in the literature as \textit{value-related dependencies}~\cite{carlshamre_industrial_2001,li_integrated_2010}, \textit{CVALUE dependencies} ~\cite{carlshamre_industrial_2001}, \textit{increases/decreases\_value\_of dependencies} ~\cite{dahlstedt2005requirements,zhang_investigating_2014}, and \textit{positive/negative value} dependencies~\cite{karlsson_improved_1997}. We use the term \textit{value-related dependencies} consistently throughout our study. 

On the other hand, the strengths of dependencies among software features vary from large to insignificant in real world software projects~\cite{carlshamre_industrial_2001,mougouei2016factoring,ngo_fuzzy_2005,wang_simulation_2012}. As such, it is important to consider not only the existence but also the strengths of value-related dependencies~\cite{dahlstedt2005requirements, brasil_multiobjective_2012,harman_exact_2014} in order to properly capture the influences of value-related dependencies on the values of software features. Carlshamre \textit{et al.} ~\cite{carlshamre_industrial_2001} observed the need to consider the strength of dependencies among software features. But, they did not go further on how to achieve this. 

Despite their importance and wide recognition~\cite{carlshamre_industrial_2001,carlshamre_release_2002,pitangueira2015software}, value-related feature dependencies have not been properly addressed in the existing software release planning models. The existing release planning models either completely ignore feature dependencies~\cite{karlsson_optimizing_1997,jung_optimizing_1998,ruhe_trade_2003} or threat those dependencies as binary relations~\cite{li_integrated_2010,sagrado_multi_objective_2013,boschetti_lagrangian_2014,brasil_multiobjective_2012,bagnall_next_2001,Greer_evolutionary_2004,baker_search_2006} without considering the strengths of those dependencies. Another group of the release planning models proposed considering value-related dependencies through direct estimation of the amount of the increased (decreased) values of software features when they are paired with other features. Such pairwise estimation however, is a complex process~\cite{carlshamre_industrial_2001} as it requires manual estimations for all pairs of features. Moreover, pairwise estimations cannot capture implicit influences of software features on the values of each other. Another problem with these models is that they do not provide any logic for estimating the amount of the increased/decreased values as they are overlooking the relationship between user preferences and value-related dependencies among features.

To tackle these issues, we have proposed mining user preferences~\cite{holland2003preference} for software features in order to discover value-related feature dependencies and their characteristics (quality and strength). In doing so, we have made four main contributions. 

First, we have presented a semi-automated approach toward mining user preferences for identification of value-related dependencies among software features and computes the qualities (positivity/negativity) and the strengths of value-related dependencies among software features using the Eells'~\cite{eells1991probabilistic} measure of casual strength. 

Second, we have made use of a resampling technique for generating samples of large (enough) quantities based on the estimated distribution of the collected data (user preferences). To achieve this, a distribution estimator is used to generate new samples based on a correlated multivariate Bernoulli distribution of collected user preferences~\cite{kroese2014statistical}. The proposed sampling technique is computationally efficient and feasible for large numbers of software features~\cite{macke2009generating}.

Third, we have demonstrated using fuzzy graphs~\cite{rosenfeld_fuzzygraph_1975} and their algebraic structure~\cite{kalampakas_fuzzy_2013} for modeling value-related dependencies. On this basis, value-related dependencies are modeled as fuzzy relations~\cite{carlshamre_industrial_2001,ngo_fuzzy_2005,ngo2005measuring,cirica_fuzzy_2010,liu_imprecise_1996} whose strengths are captured by their corresponding fuzzy membership functions.

Finally, we have formulated an integer programming model for \textit{Dependency-aware Software Release Planning (DA-SRP)} that maximizes the \textit{Overall Value} of an optimal subset of features while considering the influences of value-related dependencies extracted from user preferences. 

Validity and practicality of the work are verified through studying a real world software project. 

%

%

%% file: related.tex
\section{Related work}
\label{sec_related}

Value-related dependencies are known to be of the most common types of  dependencies among features (requirements) of software projects~\cite{carlshamre_industrial_2001}. Value-related dependencies may be inferred from intrinsic (structural/semantic) dependencies~\cite{zhang_investigating_2014} among software features. Examples of intrinsic dependencies are given in Table~\ref{table_types_intrinsic}. For instance, in the case of an intrinsic dependency of type \textit{Precede}~\cite{zhang_investigating_2014} where a feature $f_2$ precedes $f_1$, the feature $f_1$ cannot give any value if the feature $f_2$ is not selected. As such, the value of $f_1$ fully relies on the selection of $r_2$. In a similar spirit, other intrinsic dependencies such as the \textit{precondition}~\cite{k_process_centered_1996}, \textit{requires}~\cite{dahlstedt2005requirements} and \textit{conflicts}~\cite{k_process_centered_1996} dependencies also have value-related implications. 

Nonetheless, value-related dependencies may even exist in the absence of the intrinsic dependencies among software features. Such value-related dependencies have been referred to in the literature as \textit{additional value} dependencies~\cite{zhang_investigating_2014}. An example of such dependencies is a mobile application where users can listen to music while browsing photos, the feature \textit{``$f_1:$ users can listen to music"} positively influences the value of the feature \textit{``$f_2:$ users can browse photos"}. In this example there is no intrinsic dependency between $f_1$ and $f_2$ while the value of $f_2$ is influenced by $f_1$. Moreover, as explained by Carlshamre \textit{et al.}~\cite{carlshamre_industrial_2001} feature (requirement) dependencies in general are fuzzy relations~\cite{carlshamre_industrial_2001,ngo_fuzzy_2005,tang_using_2007} in the sense that the strength of those dependencies vary ~\cite{dahlstedt_moulding_2003,Robinson_RIM_2003,ngo2005measuring,ramesh_toward_2001} from large to insignificant~\cite{carlshamre_industrial_2001,wang_simulation_2012}. 

\begin{sidewaystable}
	\caption{Intrinsic feature (requirement) dependencies.}
	\label{table_types_intrinsic}
	\centering
	\input{table_types_intrinsic}
\end{sidewaystable}

Software release planning models therefore, need to consider both the qualities (positivity/negativity) and the strengths of value-related dependencies among software features. On this basis, we categorize the existing release planning models into three main groups. The first group of the release planning models referred to as the \textit{BKP} models~\cite{mougouei2016factoring} have completely ignored value-related feature dependencies~\cite{karlsson_optimizing_1997,jung_optimizing_1998,ruhe_trade_2003} by formulating release planning as a classical binary knapsack problem. Binary knapsack formulation of release planning is given in~(\ref{Eq_BKP}) where for a given set of features $F=\{f_1,...,f_n\}$, $v_i$ and $c_i$ denote the estimated value and the estimated cost of a feature $f_i$ respectively. Also $b$ specifies the available budget and $x_i$ denotes whether a feature $f_i$ is selected or otherwise.  

 \begin{align}
 \label{Eq_BKP}
  \text{Maximize} &\sum_{i=1}^{n} v_i x_i.   \\
  \label{Eq_BKP_c1}
  \text{Subject to} &\sum_{i=1}^{n} c_i x_i \leq b\\
  \label{Eq_BKP_c2}
  & x_i \in \{0,1\}
 \end{align}

The second group of the release planning models~\cite{li_integrated_2010,sagrado_multi_objective_2013,boschetti_lagrangian_2014,brasil_multiobjective_2012,bagnall_next_2001,Greer_evolutionary_2004,baker_search_2006} referred to as the \textit{BKP-PC} models~\cite{mougouei2016factoring} only allow for formulating dependencies of full strengths in terms of precedence constraints as given in~(\ref{Eq_BKP}). In this equation a positive dependency from a feature $f_j$ to a feature $f_k$ is denoted by $x_j\le x_k$ while a negative dependency from $f_j$ to $f_k$ is specified as $x_j\le 1-x_k$. Also, decision variable $x_i$ denotes whether $f_i$ is selected ($x_i=1$) or otherwise ($x_i=0$). 

As such, BKP-PC models only capture value-related dependencies of full strengths mainly the ones inferred from intrinsic dependencies (e.g. precede/requires/conflicts-with dependencies). Nevertheless, not all value-related dependencies are of full strength~\cite{mougouei2016factoring}. 

\begin{align}
\label{Eq_BKP-PC}
      &\begin{cases}
      x_j \le x_k   &\textit{ if $r_j$ positively depends on $r_k$} \\
      x_j \le 1-x_k &\textit{ if $r_j$ negatively depends on $r_k$} \\
      \end{cases}\\
      & x_i \in \{0,1\}.
\end{align}

As a result of treating all value-related dependencies as binary relations (strength of 0 or strength of 1) in BKP-PC models, ignoring (selecting) a feature $f_i$ during a release planning will result in ignoring all of the features which $f_i$ has a positive (negative) influence on even if budget is available for their implementation~\cite{li_integrated_2010}. This results in a condition which is referred to as the \textit{Selection Deficiency Problem (SDP)}~\cite{mougouei2016factoring}. As a result of the \textit{SDP}, any increase in the number of value-related dependencies (formulated as precedence constraints) would dramatically depreciate the value of an optimal subset of the features ~\cite{li_integrated_2010,mougouei2016factoring}. 

The third group of the existing release planning models~\cite{van_den_akker_flexible_2005,li_integrated_2010,sagrado_multi_objective_2013,Zhang_RIM_2013} referred to as the \textit{Increase/Decrease} models consider value-related dependencies through estimating the amount of the increased/decreased values resulted by selecting various pairs of features and imposing those increments/decrements on the values of the features as given in~(\ref{Eq_pair}). The amount of the increased/decreased values for each pair of features ($f_i,f_j$) is denoted as ($w_{ij}$) which will be achieved through estimating the value of the pair when selected together ($y_{ij}=1$). There are several problems however, with the \textit{Increase/Decrease} models as listed below. 

\begin{align}
\label{Eq_pair}
\text{Maximize }  &\sum_{i=1}^{n} v_i x_i + \sum_{i=1}^{n}\sum_{j=1}^{n} w_{i,j} y_{i,j}\\
\label{Eq_BKP-Others_c1}
\text{Subject to} &\sum_{i=1}^{n} c_i  x_i \leq b \\
\label{Eq_BKP-Others_c2}
& x_i,y_{ij} \in \{0,1\}
\end{align}

First, using pairwise estimations for identification of dependencies among software features is a complex process as it requires $\frac{n(n-1)}{2}$ (manual) estimations for a software of $n$ features. Such complexity can severely impact the practicality of pairwise comparison as the number of features grow.

Second, the increase/decrease models do not capture the relationship between the user preferences and the value-related dependencies among software features. This results in overlooking the impact of the user preferences on the values of software features during a release planning. As such, the increase/decrease models do not specify how to estimate the amounts of the increased/decreased values for pairs of features.

Third, although pairwise estimation allows for estimating the amount of the increased/decreased value of a pair, it does not specify the direction of the influence. In other words, increase/decrease models do not consider if a feature $f_i$ is increasing the value of $f_j$ or the other way round.  

Finally, pairwise estimations cannot be used to infer implicit value-related dependencies. For instance, consider the features $F=\{f_1,f_2,f_3,f_4\}$ with the estimated values of  $V=\{v_1,v_2,v_3,v_4\}$ where selecting $f_1$ and $f_2$ together increases the value of these pair to $v_1+v_2+v_{1,2}$ and the estimated value of the features $f_3$ and $f_4$ increases to $v_3+v_4+v_{3,4}$ when selected together. From these however, we can not infer any increase in the value of $f_1$ and $f_3$.

%% file: table_types_intrinsic.tex
\LARGE
\resizebox {1\textwidth }{!}{
	\begin{tabular}{|l|l|l|}
		\toprule[1.5 pt]
		\textbf{\cellcolor{black}\textcolor{white}{Dependency Type}} &
		\textbf{\cellcolor{black}\textcolor{white}{Description}} &
		\textbf{\cellcolor{black}\textcolor{white}{Example}}
		\bigstrut\\
		\hline
		\begin{tabular}[c]{@{}l@{}}Constrain~\cite{zhang_investigating_2014,k_process_centered_1996}\end{tabular} &
	    \begin{tabular}[c]{@{}l@{}}\textit{ }\textbullet\textit{ }One feature is a constraint of another feature.\end{tabular}  &
		\begin{tabular}[c]{@{}l@{}}			
			\textit{ }\textbullet\textit{ } $r_1:$ system should respond to users in 3 seconds \\
			\textit{ }\textbullet\textit{ } $r_2:$ users search books using book title or ID \\
			\textit{ }\textbullet\textit{ } $r_1$ constrains $r_2$
		\end{tabular}
		\bigstrut\\ \hline
		
		\begin{tabular}[c]{@{}l@{}}Precede~\cite{zhang_investigating_2014}, Precondition~\cite{k_process_centered_1996}, Requires~\cite{dahlstedt2005requirements}\end{tabular} &
		\begin{tabular}[c]{@{}l@{}}\textit{ }\textbullet\textit{ } If function A precedes function B\end{tabular} &
		\begin{tabular}[c]{@{}l@{}}			
			\textit{ }\textbullet\textit{ } $r_1:$ a client initiates a valuation request and the system checks the quality of the request\\
			\textit{ }\textbullet\textit{ } $r_2:$ users search books using book title or ID\\
			\textit{ }\textbullet\textit{ } $r_1$ precedes $r_2$\\
		\end{tabular}
		\bigstrut\\ \hline
		
		\begin{tabular}[c]{@{}l@{}}Be\_similar\_to~\cite{zhang_investigating_2014}, Similar~\cite{k_process_centered_1996}, Similar\_to~\cite{dahlstedt2005requirements}\end{tabular} &
		\begin{tabular}[c]{@{}l@{}}
				\textit{ }\textbullet\textit{ } (a) If two features share similar data information.\\ 
				\textit{ }\textbullet\textit{ } (b) If two features complete similar tasks.
		\end{tabular} &
		\begin{tabular}[c]{@{}l@{}}
				\textit{ }\textbullet\textit{ } (a) Adding a new book record is similar to modifying an old book as they share the book record.\\
				\textit{ }\textbullet\textit{ } (b) A property valuation could be complemented through curbside or desktop valuation.
		\end{tabular}
		\bigstrut\\ \hline
		
		\begin{tabular}[c]{@{}l@{}}Conflicts~\cite{zhang_investigating_2014,k_process_centered_1996}, Conflicts\_with~\cite{dahlstedt2005requirements}\end{tabular} &
		\begin{tabular}[c]{@{}l@{}} \textit{ }\textbullet\textit{ }One feature may negatively impact another one.\end{tabular} &
		\textit{ }\textbullet\textit{ } Security may conflicts with performance.
		\bigstrut\\ \hline
		
		\begin{tabular}[c]{@{}l@{}}Be\_exception\_of~\cite{zhang_investigating_2014}\end{tabular} &
		\begin{tabular}[c]{@{}l@{}}\textit{ }\textbullet\textit{ }One feature is an exceptional event of another one.\end{tabular} &
		\begin{tabular}[c]{@{}l@{}}
			\textit{ }\textbullet\textit{ } $r_1:$ A user inputs a null username and  system  reminds  him to correct it.\\
			\textit{ }\textbullet\textit{ } $r_2:$ A  user inputs his  username and  password  and  system checks validity of his information.\\
			\textit{ }\textbullet\textit{ } $r_1$ is an exception of $r_2$. 
		\end{tabular}
		\bigstrut\\ \hline
		
		\begin{tabular}[c]{@{}l@{}}Evolve\_into~\cite{zhang_investigating_2014}, Replaces~\cite{k_process_centered_1996}, Satisfies~\cite{k_process_centered_1996},\\ Based\_on~\cite{k_process_centered_1996}, Changes\_to~\cite{dahlstedt2005requirements}\end{tabular} &
		\begin{tabular}[c]{@{}l@{}} \textbullet\textit{ } feature $r_1$ is a new version of $r_2$. \end{tabular} &
		\begin{tabular}[c]{@{}l@{}}
			\textit{ }\textbullet\textit{ } $r_1:$ clients  use a desktop  PC  to upload  reports.\\
			\textit{ }\textbullet\textit{ } $r_2:$ clients use PDA to upload reports.\\
			\textit{ }\textbullet\textit{ } $r_1$ evolves into $r_2$. \\
		\end{tabular}
		\bigstrut\\ \hline
		
		\begin{tabular}[c]{@{}l@{}}Refines~\cite{zhang_investigating_2014,k_process_centered_1996}, Refines\_to~\cite{dahlstedt2005requirements}\end{tabular}&
		\begin{tabular}[c]{@{}l@{}} \textbullet\textit{ }One feature is refined by more specific features.\end{tabular} &
		\begin{tabular}[c]{@{}l@{}}
			\textit{ }\textbullet\textit{ } $r_1:$ a valuer submit a valuation report.\\
			\textit{ }\textbullet\textit{ } $r_2:$ a valuer submits a report through web system.\\
			\textit{ }\textbullet\textit{ } $r_1$ is refined as $r_2$. 
		\end{tabular}
		\bigstrut\\
		\bottomrule[1.5pt]
	\end{tabular}
	}%

%% file: mining.tex
\section{Mining User Preferences}
\label{sec_mining}

This section presents a semi-automated approach toward identification of value-related dependencies among software features through mining user preferences~\cite{do2016incorporating}. 

%% file: mining_gathering.tex
\subsection{Gathering User Preferences}
\label{sec_mining_gathering}

%

User preferences for software features can be gathered in different ways~\cite{villarroel2016release,leung2011probabilistic,holland2003preference,sayyad2013value} depending on the nature of a software release. For the very first release of a software, users' preferences could be gathered by conventional market research approaches such as conducting surveys or referring to the users' feedbacks or sales records of the similar software products in the market. 

For the future releases of a software, or when re-engineering of a software is of interest (e.g. for legacy systems) however, user feedbacks and sales records of the previous releases of the software might be used in combination with market research results to find user preferences. 

User preferences can be captured by a \textit{User Preference Matrix} as defined by Definition~\ref{def_pm}.

\begin{defn}
	\label{def_pm}
	\textit{Preference Matrix}. Let $F=\{f_1,...,f_n\}$ be the list of software features and $U=\{u_1,...,u_k\}$ specify the list of (potential) users of software whose preference are gathered (through survey or any of the above mentioned techniques). A user preference matrix $M_{n\times k}$ is a binary ($ 0/1 $) matrix of size $n \times k$ where $n$ and $k$ denote the number of software features and users respectively. Each element $m_{i,j}$ of $M_{n\times k}$ specifies whether a user $u_i$ would like a feature $f_j$ to be included in the next release od the software ($m_{i,j}=1$) or otherwise ($m_{i,j}=0$). 
\end{defn}

\begin{figure*}
	\begin{center}
		\includegraphics[scale=1]{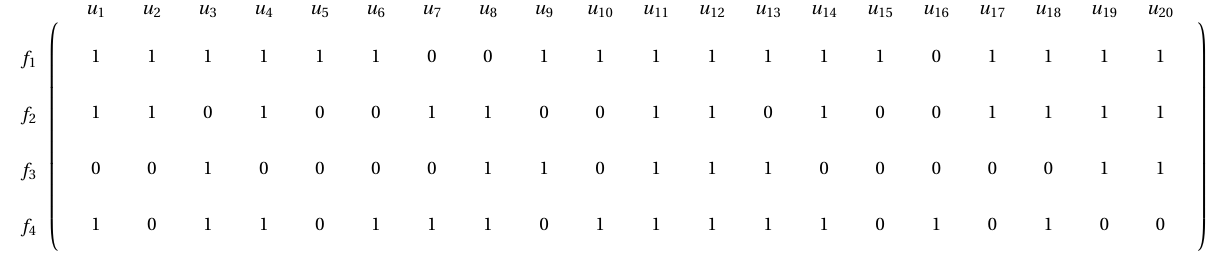}
	\end{center}
	\caption{A sample user preference matrix $M_{4\times20}$.}
	\label{fig_pm}
\end{figure*}

A sample preferences matrix $M_{4\times 20}$ is shown in Figure~\ref{fig_pm}. $m_{4,2}=0$ in $M_{4\times 20}$ specifies that the user $u_2$ does not like the feature $f_4$ to be included in the next release the software while $m_{4,3}=1$ denotes the interest of the user $u_3$ in feature $f_4$. 

%% file: mining_identification.tex
\subsection{Identification of Value-related Dependencies}
\label{sec_mining_identification}

Once user preferences are gathered, we can identify value-related dependencies and their characteristics (namely quality and strength). As value-related dependencies are naturally casual relations (selection of a feature $f_i$ may cause an increase/decrease in the value of a feature $f_j$), various measures of causal strength~\cite{sprenger2016foundations,Halpern01062015,pearl2009causality,janzing2013quantifying,eells1991probabilistic} can be employed for estimating the strengths and qualities (positivity/negativity) of value-related dependencies.

One of the most widely adopted measures of causal strength is Eells' measure of causal strength~\cite{eells1991probabilistic}, dented by $\eta_{i,j}$ as given in (\ref{Eq_Eells}). $\eta_{i,j}$ in (\ref{Eq_Eells}) specifies the casual strength of a value-related dependency from a feature $f_i$ to $f_j$ where selecting or ignoring $f_j$ causes an increase or decrease on the value of $f_i$. Eells' measure of (causal) strength properly captures both positive and negative value-related dependencies between a pair of features through subtracting the conditional probability $p(f_i|\bar{f_j})$ from $p(f_i|f_j)$ where conditional probabilities $p(f_i|\bar{f_j})$ and $p(f_i|f_j)$ denote strengths of positive and negative dependencies from $f_i$ to $f_j$ respectively. 

\begin{align}
\label{Eq_Eells}
& \eta_{i,j}= p(f_i|f_j) - p(f_i|\bar{f_j}) ,\phantom{s}\eta_{i,j} \in [-1,1]
\end{align}

\begin{figure*}[!htb]
	\begin{center}
		\subfigure[$P_{4 \times 4}$]{%
			\label{fig_eta_p}
			\includegraphics[scale=0.3]{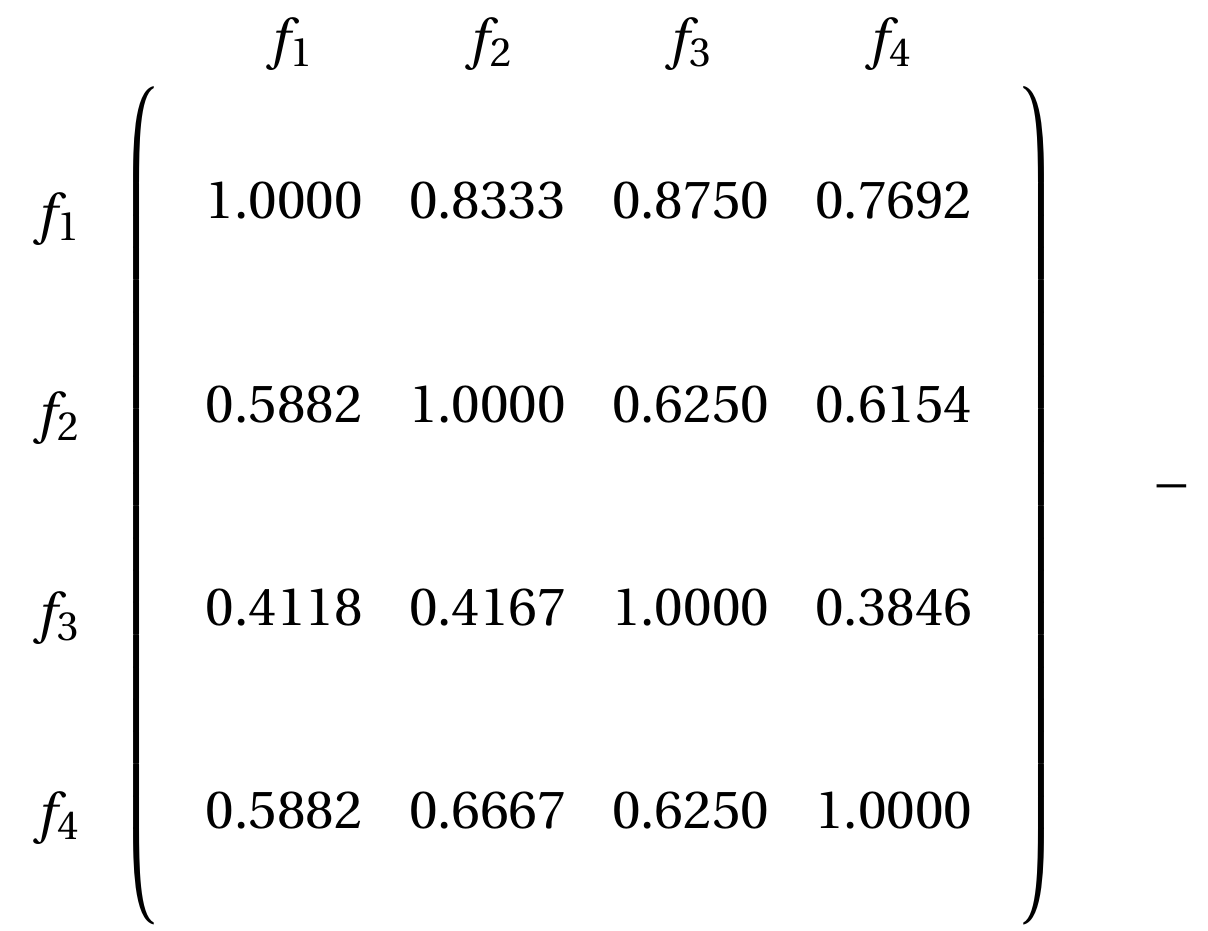}
		}
		\subfigure[$\bar{P}_{4 \times 4}$]{%
			\label{fig_eta_p_bar}
			\includegraphics[scale=0.3]{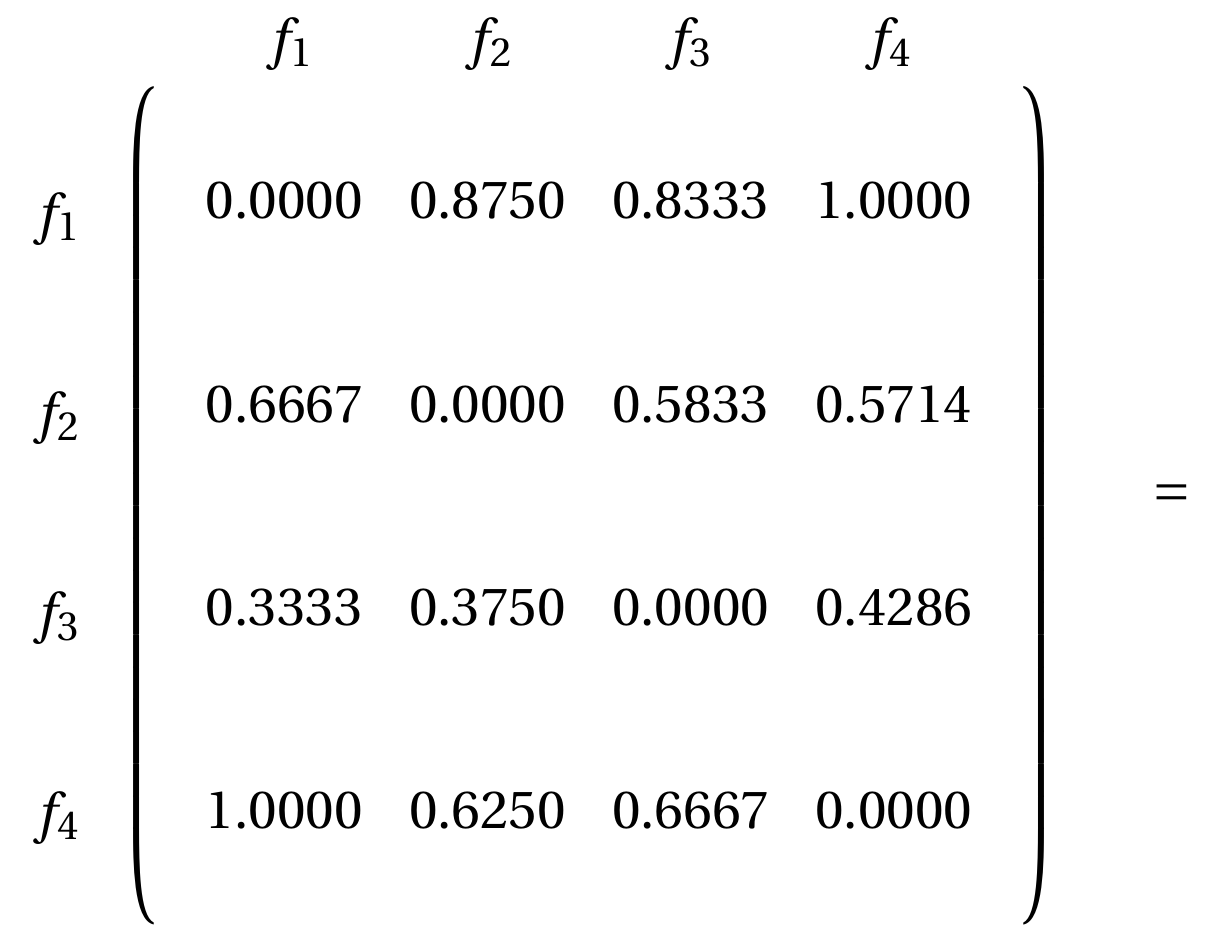}
		}
		\subfigure[$\bm{\eta}_{4 \times 4}$]{%
			\label{fig_eta_eta}
			\includegraphics[scale=0.3]{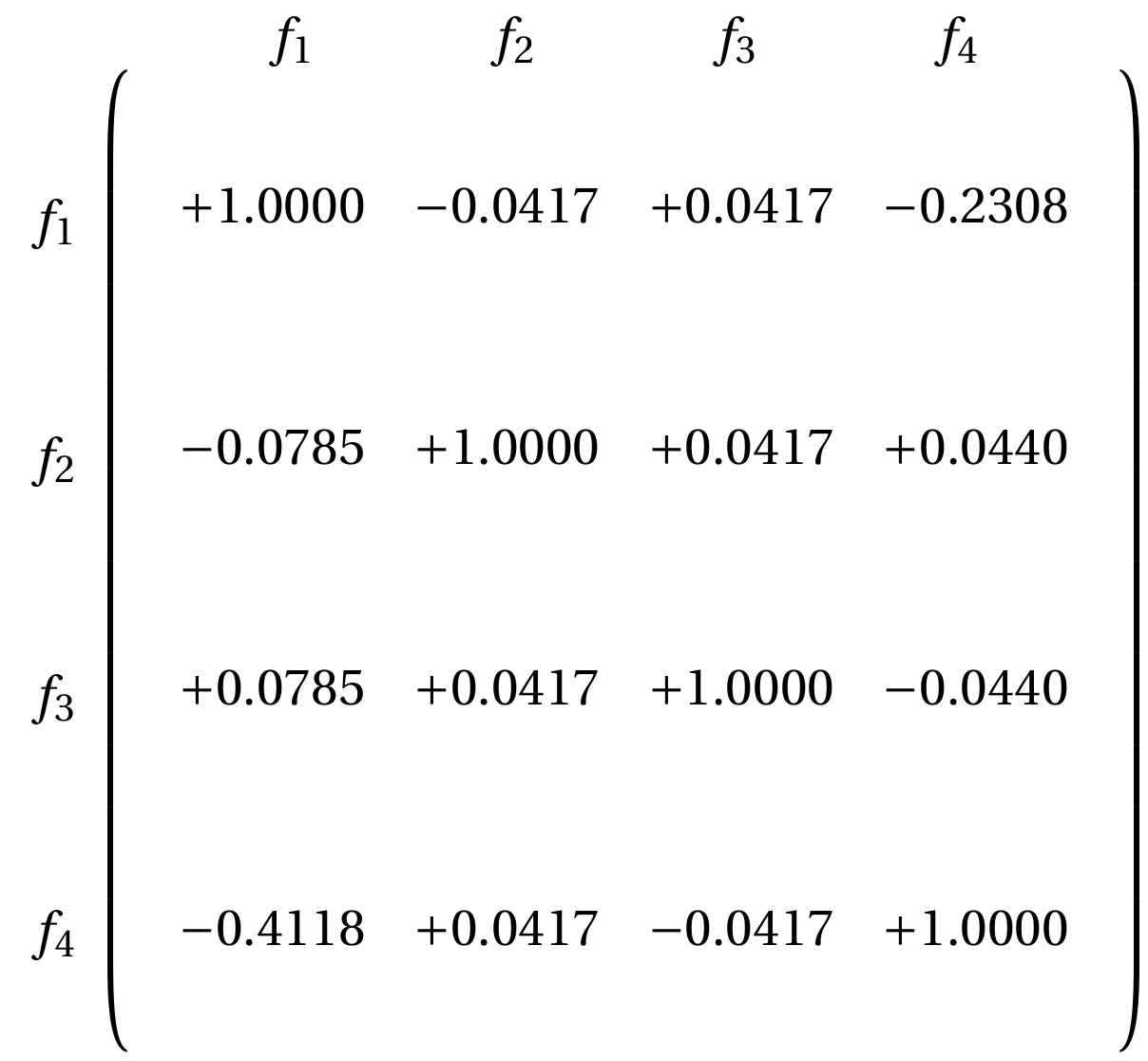}
		}
	\end{center}
	\caption{%
		Computing the Eells measure of casual strength for preference matrix $M_{4\times20}$ of Figure~\ref{fig_pm}.
	}%
	\label{fig_eta}
\end{figure*}

Matrix $P_{4\times4}$ (Figures~\ref{fig_eta_p}) and Matrix $\bar{P}_{4\times4}$ (Figure~\ref{fig_eta_p_bar}) demonstrate strengths of positive and negative dependencies between pairs of features in the preference matrix $M_{4\times 8}$ of Figure~\ref{fig_pm}. An element $p_{i,j}$ of matrix $P_{4\times4}$ denotes the strength of a positive dependence from $f_i$ to $f_j$ while an element $\bar{p}_{i,j}$ of $\bar{P}_{4\times4}$ gives the strength of a negative dependency from $f_i$ to $f_j$. As such, subtracting each element $\bar{p}_{i,j}$ of $\bar{P}_{4 \times 4}$ from its corresponding element $p_{i,j}$ in $P_{4 \times 4}$ will give the Eells' casual strength for a dependency from $f_i$ to $f_j$. 

Algorithm~\ref{alg_identification} gives the steps for computing the measure of casual strength for a given preference matrix $M_{n\times k}$. In this algorithm, an element $\lambda_{i,j}$ in matrix $\lambda_{n\times 2n}$ denotes the number of times that a features $f_i$ is selected with a feature $f_j$ in pair. Also, an element $\lambda_{i,j+n}$ gives the number of times that users have selected $f_i$ without selecting $f_j$. It is clear that, $\lambda_{i,i}$ gives the number of occurrences of $f_i$ in $M_{n\times k}$ while $\lambda_{i,i+n}=0$.  

\begin{algorithm}[!htb]
	\small
	\caption{Computing Eells' measure of strength.}
	\label{alg_identification}
		\begin{algorithmic}[1]
			\REQUIRE \textit{users' preference matrix $M_{n\times k}$} 
			\ENSURE  \textit{Matrix $\bm{\eta}_{n \times n}$}
			\STATE $\bm{\eta}_{n \times n} \leftarrow 0$ 
			\STATE $\bm{\lambda}{n \times 2n} \leftarrow 0$ 

			\FOR{\textbf{each} $u_{t} \in U$}
	     		\FOR{\textbf{each} $f_{i} \in F$}
					\FOR{\textbf{each} $f_{j} \in F$}
						\IF{$m_{i,t}=1$}
							\IF{$m_{j,t}=1$}
								\STATE $\lambda_{i,j} \leftarrow \lambda_{i,j} + 1$ 
							\ELSE
								\STATE $\lambda_{i,j+n} \leftarrow \lambda_{i,j+n} + 1$ 
							\ENDIF
						\ENDIF
					\ENDFOR
				\ENDFOR
			\ENDFOR
		    
		    \FOR{\textbf{each} $f_{i} \in F$}
				\FOR{\textbf{each} $f_{j} \in F$}
					\STATE $\eta_{i,j} \leftarrow \frac{\lambda_{i,j}}{\lambda_{j,j}} - \frac{\lambda_{i,j+n}}{\lambda_{j+n,j+n}}$ 
				\ENDFOR
			\ENDFOR
		\end{algorithmic}
\end{algorithm}

%% file: mining_resampling.tex
\subsection{Resampling}
\label{sec_mining_resampling}

\begin{figure}[!htb]
	\centering
	\centerline{\includegraphics[scale=0.81]{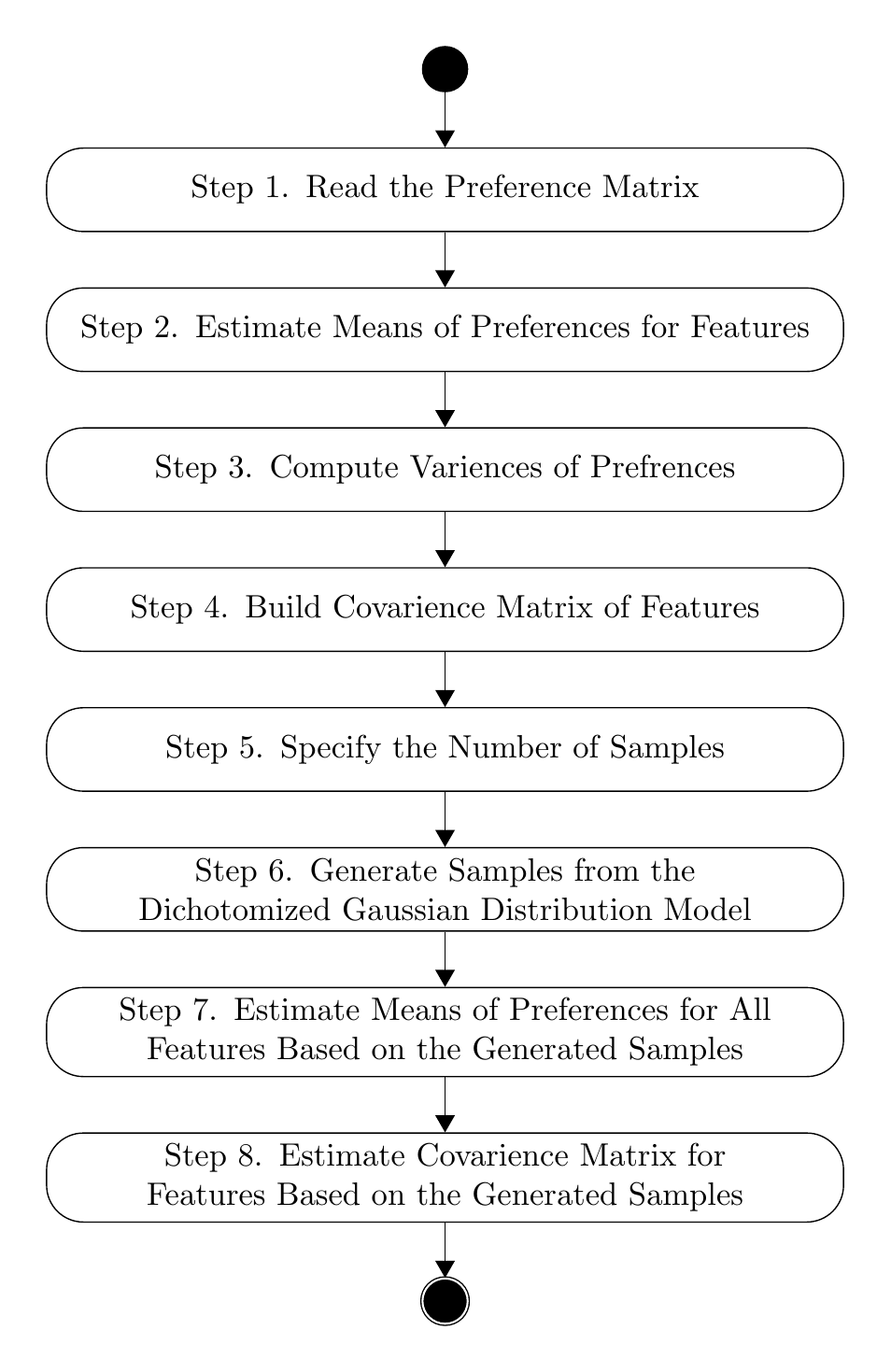}}
	\caption{Steps for generating samples from user preferences.}
	\label{fig_resampling}
\end{figure}

When collecting user preferences in large quantities is not feasible (e.g. only a small number of users are familiar with the features of the software), statistical methods such as resampling can be used to generate samples of large (enough) quantities based on the estimated distribution of the collected data (user preferences). To achieve this, a distribution estimator must be used to generate new samples based on a correlated multivariate Bernoulli (select or not select) distribution of collected user preferences. Hence, we have made use of an interesting approach introduced by Macke \textit{et al.}~\cite{macke2009generating} in the field of Bioinformatics which was initially used for generating samples of artificial spike trains (between pair of neurons) with specified correlation structures. 

Using the Macke's approach, we generate larger samples of user preferences using a Latent Multivariate Gaussian model~\cite{kroese2014statistical}. The process as given in Figure~\ref{fig_resampling} starts with reading preference matrix of users (Step 1) and continues with estimating the means of user preferences (Step 2) and computing the variances of user preferences (Step 3) for each feature. Then, covariances matrix of the features will be calculated (Step 4) for generating new samples. After that, number of samples will be specified (Step 5) and samples  will be generated based on the Dichotomized Gaussian Distribution model discussed in~\cite{kroese2014statistical} (Step 6). We then, estimate means of preferences for features in the generated samples (Step 7) and finally estimate the covariance matrix of the features (Step 8). Steps 7,8 are performed to evaluate the precision of the process by comparing the means and covariance matrix of the generated samples against those of the initial samples directly gathered from users.    

Sampling user preferences using the approach proposed by Macke \textit{et al.}~\cite{macke2009generating} is computationally efficient and specially feasible even for large numbers of software features. Macke \textit{et al.}~\cite{macke2009generating} showed that the entropy of their model is near theoretical maximum for a wide range of parameters.


%% file: modeling.tex
\section{Modeling Value-related Dependencies}
\label{sec_modeling}

%

Using fuzzy graphs for modeling value-related feature dependencies can contribute to more accurate~\cite{} release planning as fuzzy graphs properly capture uncertainties associated with software features~\cite{mougouei2015partial,mougouei2013fuzzyBased,mougouei2014visibility,mougouei2012evaluating,mougouei2012measuring} and value-related dependencies among them. Hence, we have presented a modified version of fuzzy graphs referred to as the \textit{Feature Dependency Graph (FDG)} in Definition~\ref{def_vdg} that allows for modeling value-related dependencies and their characteristics.   

\begin{defn}
	\label{def_vdg}
	\textit{Feature Dependency Graph (FDG)}. A \textit{FDG} is a signed directed fuzzy graph~\cite{Wasserman1994} $G=(F,\sigma,\rho)$ in which a non-empty set of features $F=\{f_1,...,f_n\}$ constitute the graph nodes. The qualitative function $\sigma: F\times F\rightarrow \{+,-,\pm\}$ denotes the qualities: positive ($ + $), negative ($ - $), and unspecified ($\pm$) of value-related dependencies. Also, the membership function $\rho: F\times F\rightarrow [0,1]$ denotes the strengths of value-related dependencies (edges of the graph). As such, a pair of features $(f_i,f_j)$ with $\rho_{i,j}\neq 0$ denotes a value-related dependency from $f_i$ to $f_j$ stating that the value of $f_i$ depends on $f_j$. It is clear that we have $\rho_{i,j}=0$ and $\sigma_{i,j}=\pm$ if the value of a feature $f_i$ is not influenced by $f_j$. 
\end{defn}

\begin{figure}[!htb]
	\begin{center}
		\label{fig_ex_selection}
		\includegraphics[scale=0.6]{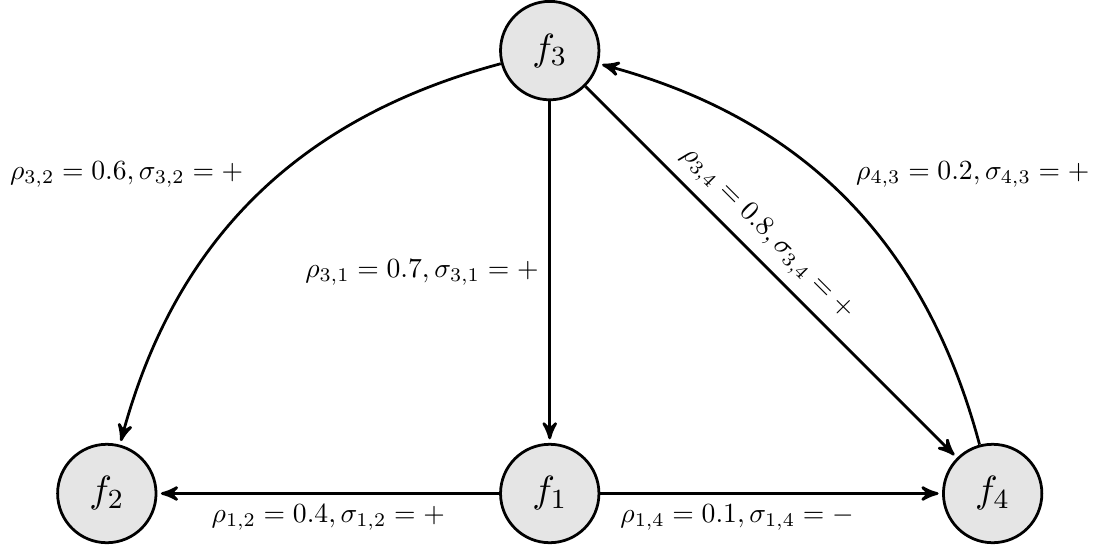}
	\end{center}
	\caption{%
		FDG of Example~\ref{ex_fuzzy_relation}.}%
	\label{fig_ex_vdg}
\end{figure}

\begin{exmp}
	\label{ex_fuzzy_relation}
	Consider the \textit{FDG} $G=(F,\sigma,\rho)$ with features $F=\{f_1,f_2,f_3,f_4\}$. As given by Figure~\ref{fig_ex_vdg} qualities and strengths of value-related dependencies in $G$ are specified by the function $\sigma$ and the membership function $\rho$ respectively. For instance, $\sigma_{1,2}=+$ and $\rho_{1,2}=0.4$ state that selection of feature $f_2$ has a positive influence on the value of feature $f_1$ and the strength of this influence is $0.4$. In a similar way, $\sigma_{1,4}=-$ and $\rho_{1,4}=0.1$ state that selection of feature $f_4$ has a negative influence on the value of $f_1$ and the strength of this influence is $0.1$. 
\end{exmp}

\begin{figure}[!htb]
	\begin{center}
		\subfigure[$$]{%
			\label{fig_membership_1}
			\includegraphics[scale=0.71]{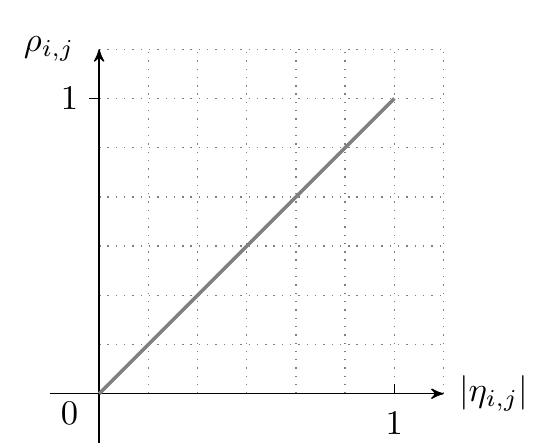}
		}
		\subfigure[$$]{%
			\label{fig_membership_2}
			\includegraphics[scale=0.71]{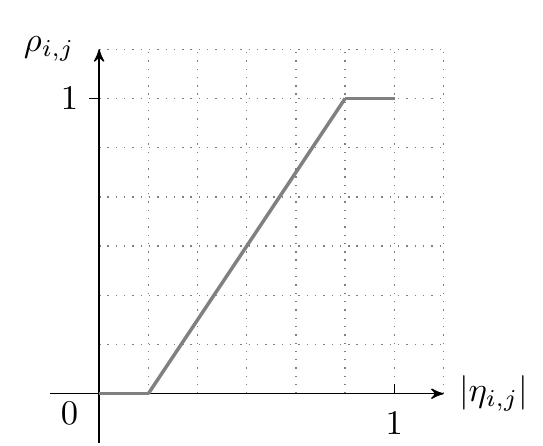}
		}
		\subfigure[$$]{%
			\label{fig_membership_3}
			\includegraphics[scale=0.71]{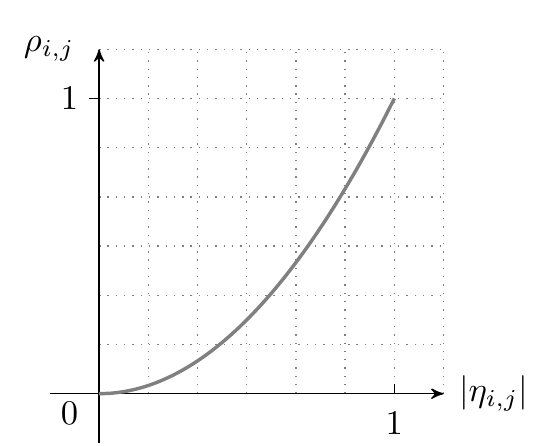}
		}
		\subfigure[$$]{%
			\label{fig_membership_4}
			\includegraphics[scale=0.71]{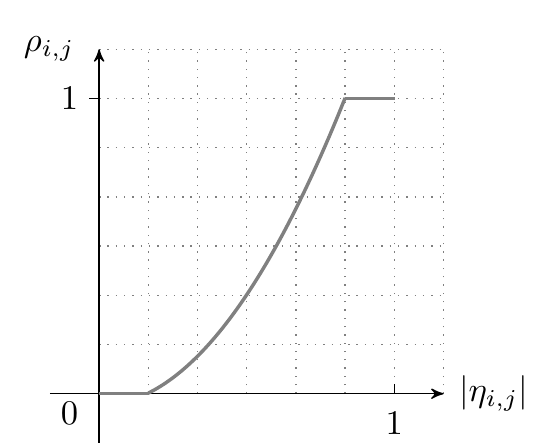}
		}
	\end{center}

	\caption{%
		Sample membership functions $\rho_{i,j}$.
	}%
	\label{fig_membership}
\end{figure}

To compute the strength of a value-related dependency $d=(f_i,f_j)$ in a feature dependency graph $G=(F,\sigma,\rho)$, (\ref{Eq_strengthMeasure}) gives a mapping from $\eta_{i,j}$ to the fuzzy membership function of the strength of $d$ specified by $\rho: F\times F\rightarrow [0,1]$. This mapping is demonstrated in Figure~\ref{fig_membership_1}. In a similar spirit, (\ref{Eq_qualityMeasure}) gives a mapping from $\eta_{i,j}$ to the qualitative function $\sigma: F\times F\rightarrow \{+,-,\pm\}$ in order to specify the quality of the value-related dependency $d$. 

\begin{align}
\label{Eq_strengthMeasure}
& \rho_{i,j}= |\eta_{i,j}|\\
\label{Eq_qualityMeasure}
& \sigma_{i,j} =  \begin{cases}
+ & \text{if }\phantom{s}  \eta_{i,j} > 0 \\
- & \text{if }\phantom{s}  \eta_{i,j} < 0 \\
\pm & \text{if }\phantom{s} \eta_{i,j} = 0 \\
\end{cases}
\end{align}

\begin{figure*}[!htb]
	\begin{center}
		\subfigure[Strength Matrix $(\bm{\rho}_{4\times4})$]{%
			\label{fig_rhosigmaprecede_rho}
			\includegraphics[scale=0.3]{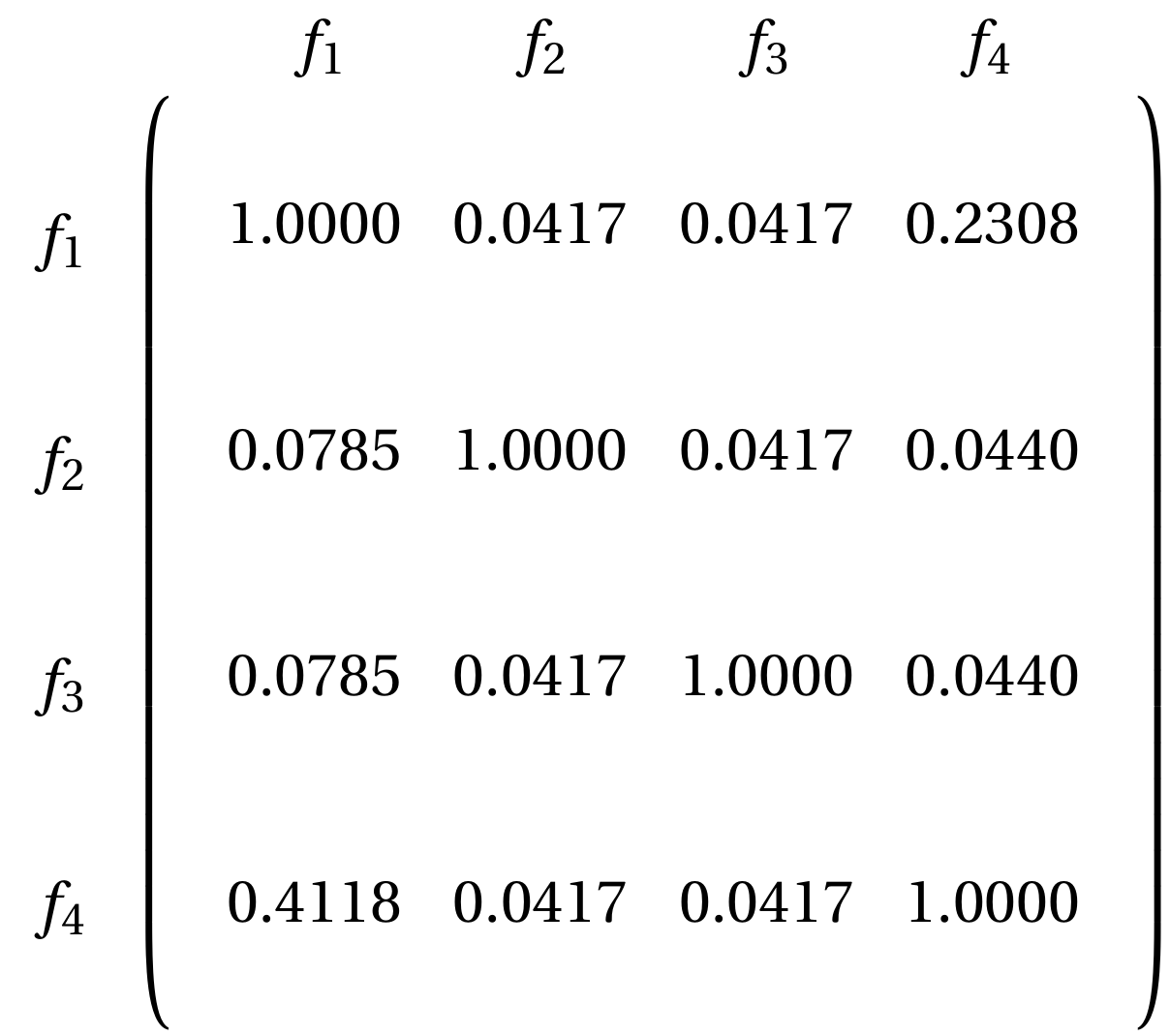}
		}
		\subfigure[Quality Matrix $(\bm{\sigma}_{4\times4})$]{%
			\label{fig_rhosigmaprecede_sigma}
			\includegraphics[scale=0.3]{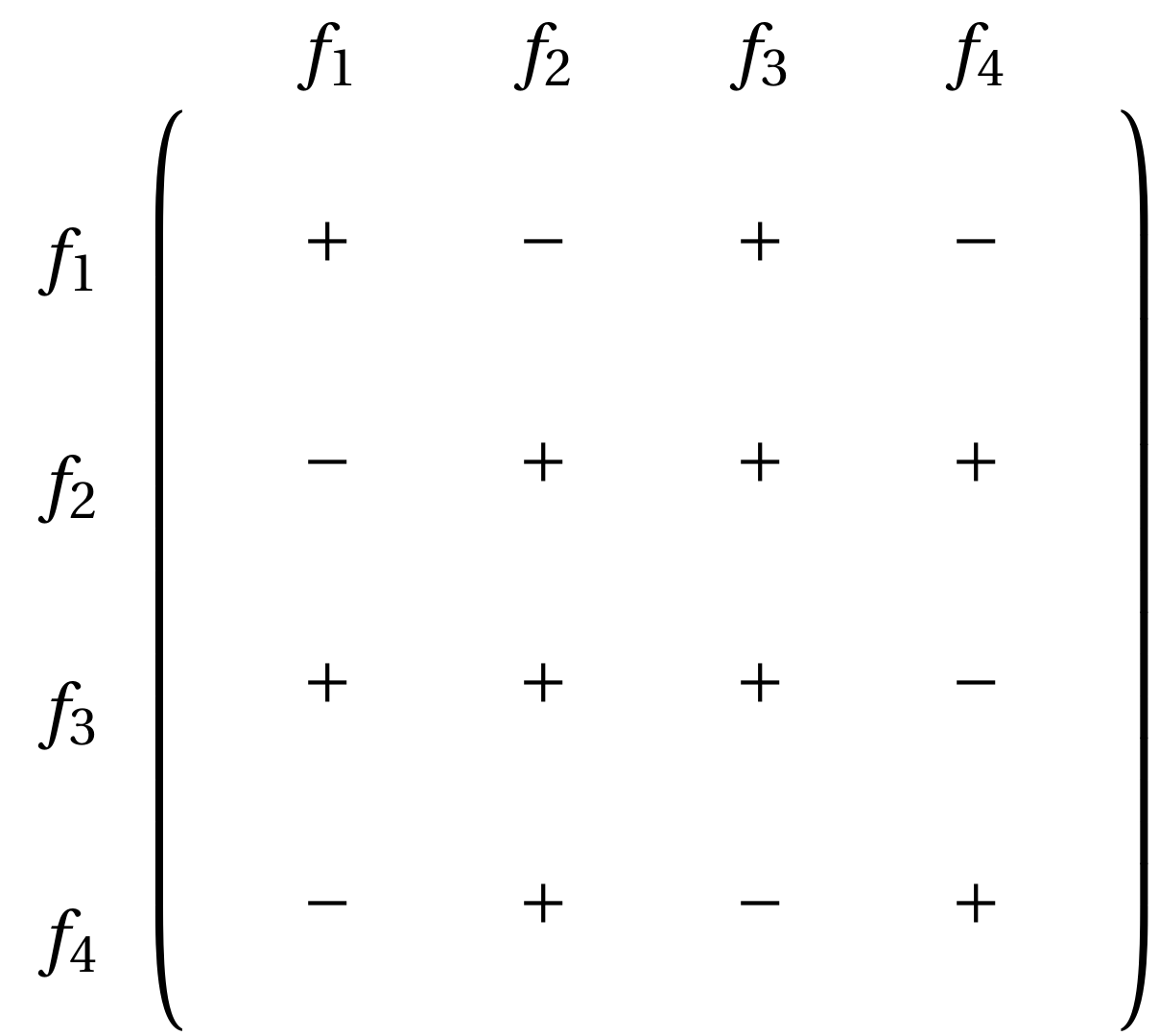}
		}
		\subfigure[Precedence Matrix $(\bm{\gamma}_{4\times4})$]{%
			\label{fig_rhosigmaprecede_precede}
			\includegraphics[scale=0.3]{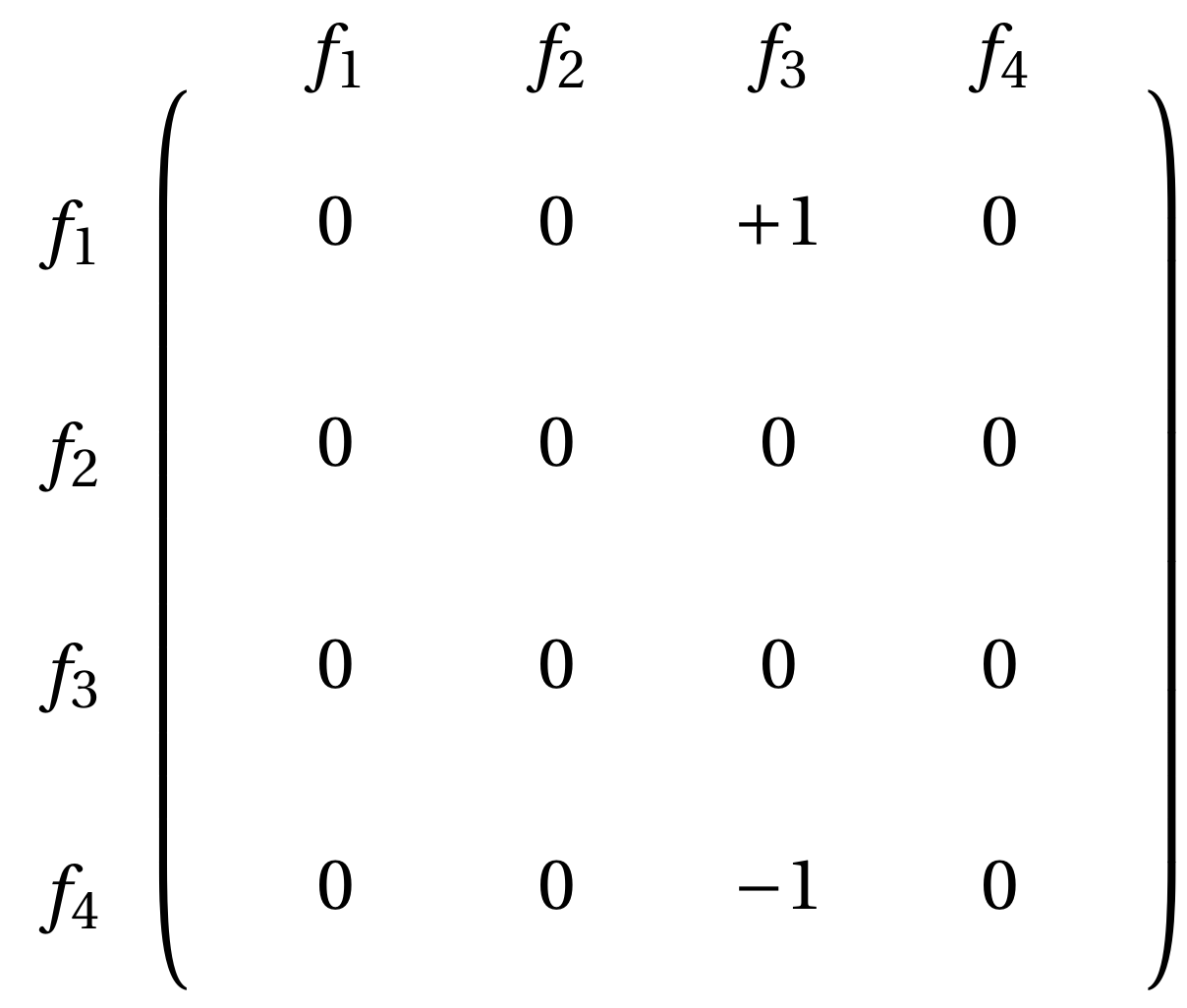}
		}
	\end{center}
	\caption{%
		Strength, Quality, and Precedence Matrices for preference matrix of Figure~\ref{fig_pm}.
	}%
	\label{fig_rhosigmaprecede}
\end{figure*}

A positive $\eta_{i,j}$ indicates that the strength of the positive dependency from $f_i$ to $f_j$ is greater than the strength of its corresponding negative dependency: $p(f_i|f_j) > p(f_i|\bar{f_j})$ and therefore the quality of $d$ would be positive ($\sigma_{i,j}=+$). Similarly, a negative $\eta_{i,j}$ indicates $p(f_i|\bar{f_j}) > p(f_i|f_j)$ which gives $\sigma_{i,j}=-$. Also, $p(f_i|f_j) - p(f_i|\bar{f_j})=0$ denotes that the quality of the zero-strength value-related dependency $d=(f_i,f_j)$ is non-specified ($\sigma_{i,j}=\pm$). $\rho_{i,j}$ and $\sigma_{i,j}$ are computed for all pairs of features in preference matrix $M_{4\times20}$ of Figure~\ref{fig_pm} as demonstrated in Figure~\ref{fig_rhosigmaprecede_rho} and Figure~\ref{fig_rhosigmaprecede_sigma} respectively.

Mapping value-related dependencies to $\rho$ in FDG $G=(F,\sigma,\rho)$ however, does not have to be restricted to membership function of the Figure~\ref{fig_membership_1}. Other membership functions such as the one in Figure~\ref{fig_membership_2} could also be used depending on the characteristics of the user preferences and the behavior of the release planning models used. For instance, the membership function of Figure~\ref{fig_membership_1} considers value-related dependencies with casual strengths below $0.16$ ($\eta_{i,j} < 0.16$) to be too weak to be considered  while value-related dependencies with $\eta_{i,j} \geq 0.83$ are considered strong enough to be interpreted as full dependencies of full strength ($\rho_{i,j}=1$). 

Similar membership functions to that in Figure~\ref{fig_membership_1} could be particularly useful in release planning models that formulate value-related dependencies as precedence constraints (BKP-PC models). When such models used, it might be more reasonable to consider a strong value-related dependency (e.g.  $\eta_i,j \geq 0.95$) as a precedence relation (BKP-PC models only capture precedence relations which are of full strength of $\eta_{i,j}=1$) rather than ignoring it.

Figure~\ref{fig_membership_3} and Figure~\ref{fig_membership_4} depict other alternative membership functions which do not assume linearity between $|\eta_{i,j}|$ and $\rho_{i,j}$. 

As mentioned earlier, \textit{Intrinsic} dependencies of full strength among features (e.g. \textit{precede}~\cite{zhang_investigating_2014}, \textit{precondition}~\cite{k_process_centered_1996}, \textit{requires}~\cite{dahlstedt2005requirements} and \textit{conflicts}~\cite{k_process_centered_1996}) also have value implications which need to be considered while identification of value-related dependencies. Intrinsic dependencies as given in Table~\ref{table_types_intrinsic} reflect structural (semantic) dependencies among features. In the practice however, intrinsic dependencies may imply value-related dependencies among software features.

Such value-related dependencies may in some cases be captured by mining user preferences while others can hardly if ever be identified by solely relying on user preferences. For instance, a \textit{precede} dependence from a feature $f_i$ to a feature $f_j$ implies that $f_i$ can not be selected (implemented) unless $f_j$ is selected. A value implication of this would be $f_i$ can not give any value unless $f_j$ is selected. In other words, the value of $f_i$ fully relies on selection of $f_j$ ($\rho_{i,j}=1$). This value-related dependency may not be captured by mining user preferences as users may not be aware of the intrinsic dependency from $f_i$ to $f_j$. After all, users are not expected to consider such dependencies.   

Hence, it is important to be aware of value implications of intrinsic dependencies and consider them in software release planning. This could be achieved by carefully studying the structure (semantic) of a software. We specially focus on considering value-related implications of intrinsic dependencies of full strengths (\textit{precede}~\cite{zhang_investigating_2014}, \textit{precondition}~\cite{k_process_centered_1996}, \textit{requires}~\cite{dahlstedt2005requirements} and \textit{conflicts}~\cite{k_process_centered_1996}) formulated as precedence constraints. 

In doing so, a binary matrix of precede relations referred to as the \textit{precedence matrix} needs to be constructed by stakeholders of the software to capture value implications of precede relations among features. The precedence matrix $\bm{\gamma}_{4\times 4}$ for preference matrix $M_{4\times 8}$ of Figure~\ref{fig_pm} is demonstrated in Figure~\ref{fig_rhosigmaprecede_precede}. $\gamma_{1,3}=+1$ in $\bm{\gamma}_{4\times 4}$ denotes that feature $f_3$ precedes $f_1$ for structural or semantic reasons such as $f_3$ is required by $f_1$, or $f_3$ is a precondition of $f_1$. $\gamma_{4,3}=-1$ on the other hand, specifies that $f_4$ cannot be selected when $f_3$ is selected as $f_4$ conflicts with $f_3$. Examples for such relations are provided in Table~\ref{table_types_intrinsic}. Finally, we have $\gamma_{i,j}=0$ when there is no dependency of type precedence from $f_i$ to $f_j$.

\begin{figure}[!htb]
	\begin{center}
		\subfigure[Strength Matrix $(\bm{\rho}_{4\times4})$]{%
			\label{fig_rhosigma_updated_rho}
			\includegraphics[scale=0.3]{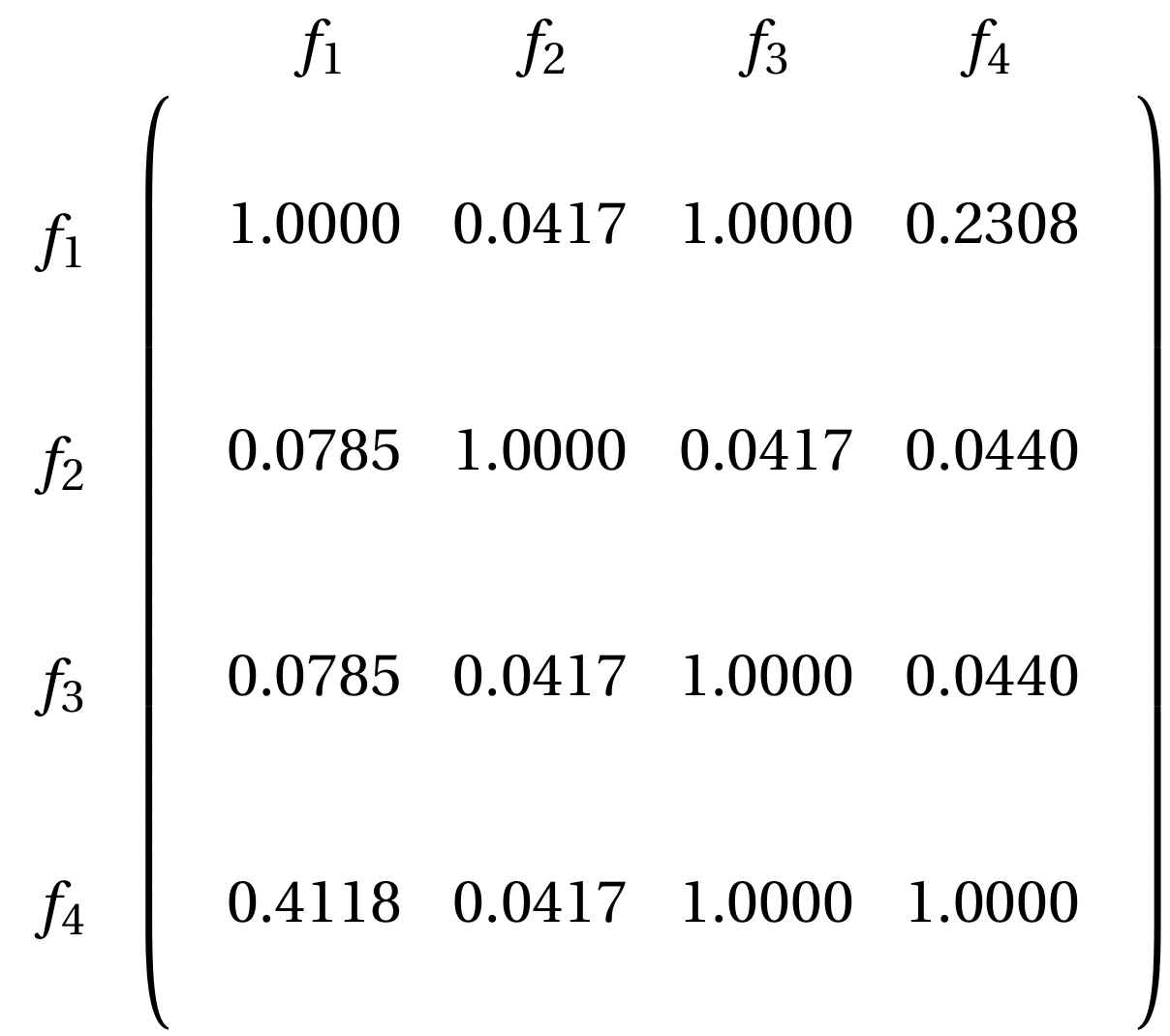}
		}
		\subfigure[Quality Matrix $(\bm{\rho}_{4\times4})$]{%
			\label{fig_rhosigma_updated_sigma}
			\includegraphics[scale=0.3]{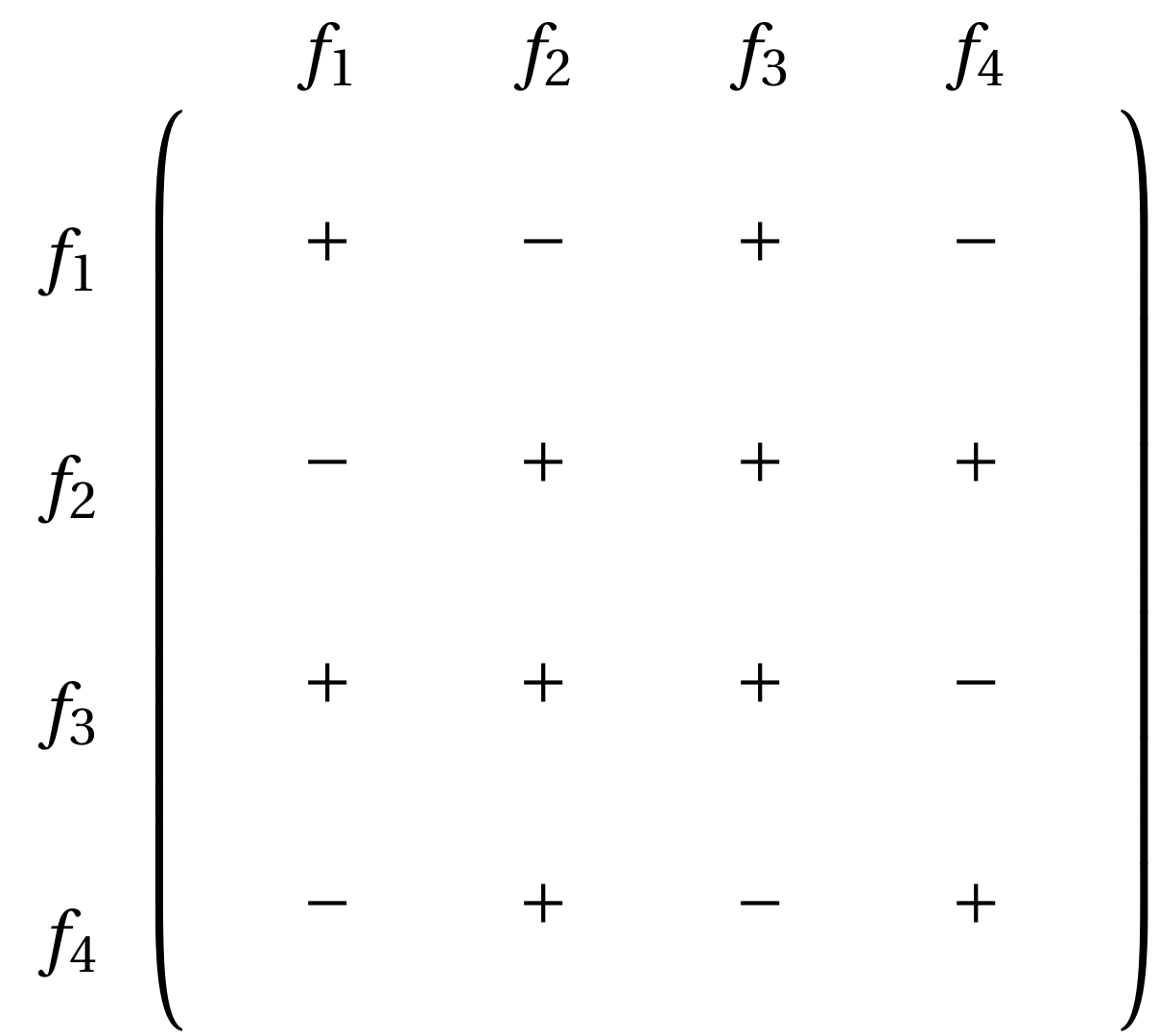}
		}
	\end{center}
	\caption{%
		Strength and Quality Matrices of Figures~\ref{fig_rhosigmaprecede_rho} and \ref{fig_rhosigmaprecede_sigma} updated based on precedence matrix of Figure~\ref{fig_rhosigmaprecede_precede}.
	}%
	\label{fig_rhosigma_updated}
\end{figure}

Strengths and qualities of value-related dependencies then, can be updated based on the precedence matrix of a software. For each element $\gamma_{i,j}\neq 0$ of the precede matrix, $\rho_{i,j}$ is set to $1$ and $\sigma_{i,j}$ is set to the sign of the precede dependence $\gamma_{i,j}$. The reason is that value implications of full dependencies extracted from structure (semantic) of the software are naturally of higher importance. Strength and quality matrices in Figure~\ref{fig_rhosigmaprecede} are updated based on the precedence matrix of Figure~\ref{fig_rhosigmaprecede_precede} as depicted in Figure~\ref{fig_rhosigma_updated}. $\rho_{1,3}=0.0417$ is updated to $\rho_{1,3}=1$ since $\gamma_{1,3}=+1$ and $\rho_{4,3}=0.0417$ is updated to $\rho{4,3}=1$ since $\gamma_{4,3}=-1$. The quality matrix $\bm{\sigma_{4 \times 4}}$ however remained unchanged. 

%% file: factoring.tex
\section{Dependency-Aware Release Planning}
\label{sec_factoring}

Now that we have identified value-related dependencies among software features from user preferences, it is time to consider them in software release. Such release planning which accounts for value-related dependencies among software features, is referred to as dependency-aware software release planning (DA-SRP). This section explains how dependency-aware software release planning can be achieved. 

As explained earlier, the value of a software feature is determined by user preferences for that feature~\cite{zhang_investigating_2014,biffl_value_2006,racheva2010business}. In other words, the higher the ratio of the users that prefer (select/use) a software feature $f_i$, the higher the expected value of $f_i$ would be. That means $f_i$ can achieve its highest expected value (which equals to its estimated value) when all users of $f_i$ are satisfied with that feature. However, this only happens when user preferences for the feature $f_i$ is not negatively influenced by selecting or ignoring other features. In other cases, a reduction from the estimated value of $f_i$ is predictable. The extent of such reduction is referred to as the penalty~\cite{wiegers_software_2009} of $f_i$ and denoted as $p_i$. 

We have made use of the algebraic structure of fuzzy graphs for modeling value-related dependencies and their characteristics (quality and strength) as explained in Section~\ref{sec_modeling}. Hence, $p_i$ is calculated by using the fuzzy OR operator (taking supremum) over the strengths of all ignored positive dependencies and selected negative dependencies of $f_i$ in its corresponding feature dependency graph. This is given in (\ref{Eq_dars_c2}) where $n$ denotes the total number of the features and $x_j$ specifies whether a feature $f_j$ is selected ($x_j=1$) or otherwise ($x_j=0$). Also, $\sigma_{i,j}\rho_{i,j}$ denotes the quality ($\sigma_{i,j}$) and the strength ($\rho_{i,j}$) of the overall influence of a feature $f_j$ on the value of $f_i$ through various dependency paths from $f_i$ to $f_j$ in a feature dependency graph. 

\vspace{1em} 

Equation (\ref{Eq_value}) derives the expected value of a software feature $f_i$ denoted by $v^\prime_i$ which captures the impact of user preferences for $f_i$ ($\phi_i$) on the value of $f_i$. $\phi_i$ is derived by subtracting the penalty $p_i$ from the ideal user preference (satisfaction) level $1$ (when every user selects or uses $f_i$). As such, the overall value (OV) of an optimal subset of features can be calculated by accumulating the expected value of the selected features as given by (\ref{Eq_ocv}) where $x_i$ denotes whether $f_i$ is selected ($x_i=1$) or otherwise ($x_i=0$). 

\begin{align}
\label{Eq_value}
& v^\prime_i = \phi_i v_i = (1-p_i) v_i  \\
\label{Eq_ocv}
&OV = \sum_{i=1}^{n} x_i \phi_i v_i, \textit{ } x_i \in \{0,1\}
\end{align}
\vspace{0.1em}

Our proposed formulation of dependency-aware software release planning, maximizes the overall value (OV) of an optimal subset of features as given by (\ref{Eq_dars})-(\ref{Eq_dars_c3}) where $x_i$ is a selection variable denoting whether a features $f_i$ is selected ($x_i=1$) or otherwise ($x_i=0$).

\begin{align}
\label{Eq_dars}
& \text{Maximize } \sum_{i=1}^{n} x_i \phi_i v_i\\
\label{Eq_dars_c1}
& \text{Subject to} \sum_{i=1}^{n} c_i x_i \leq b\\ 
\label{Eq_dars_c2}
& p_i \ge \displaystyle \bigg(\frac{\lvert \sigma_{i,j}\rho_{i,j} \rvert + (1-2x_j)\sigma_{i,j}\rho_{i,j}}{2}\bigg), i,j = 1,...,n\\
\label{Eq_dars_c3}
&\text{ }x_i \in \{0,1\}
\end{align}
\vspace{0.1em}

%% file: case.tex
\subsection{Case Study}
\label{sec_validation_case}

To demonstrate the validity and practicality of our work, we performed release planning for reengineering~\cite{Arnold_1993_SR_530234,Miller_1998_RLS_288764} an industrial software referred to as the PMS. The first version of the software (PMS-I) was introduced with $23$ features and a few bug-fixing versions of the software were released later. Stakeholders of the PMS-I recently decided to re-engineer and develop a new version of the software (PMS-II) to cope with market demands. In this regard, we designed a case study to assist stakeholders find an optimal subset of features with the highest overall value (considering users' preferences) while respecting the available budget.  

In doing so, we carried out a survey on users of earlier versions of the software to gather their preferences of the features. For each of the $23$ features of the PMS-I users where asked whether they would like that feature to remain in the new version of the software or otherwise. Users were also encouraged to propose their desired features that have not existed in the earlier versions. As a result of this, a total number of $176$ records of user preference were gathered. In total, $7$ new features were proposed by the users from which $4$ were found to be technically feasible to implement. As such, preferences of users for a total number of $27$ features were gathered to be used for release planning.

Preference matrix of the PMS-II was then constructed based on the survey results. Then, $10^6$ new samples of user preferences were generated from the gathered user preferences based on the resampling technique presented in Section~\ref{sec_mining_resampling}. A new preference matrix was then constructed based on the generated samples. Afterwards, Eells' measure of casual strength was computed for the new preference matrix of features of PMS-II using Algorithm~\ref{alg_identification} to identify potential value-related dependencies among pairs of features. Strengths (strength matrix) and qualities (quality matrix) of value-related dependencies then, were computed based on the fuzzy membership function of Figure~\ref{fig_membership_1} and~(\ref{Eq_strengthMeasure})-(\ref{Eq_qualityMeasure}). 

To specify the costs and value of the features, $5$ different stakeholders estimated the cost (value) of each feature $f_i$ of PMS-II. Then median of these $5$ estimated costs (values) was computed and scaled into the range of $[0,...,20]$ to determine the cost/value of $f_i$. 

As explained in Section~\ref{sec_mining_identification}, intrinsic dependencies of full strength among features also have value implications which need to be considered while identification of value-related dependencies. In order to capture such value-related dependencies stakeholders of the PMS-II updated the precedence matrix of the PMS-II based on the approach given in Section~\ref{sec_mining_identification}. Value-related dependencies identified from intrinsic dependencies among features of the PMS-II are highlighted in Table~\ref{table_rho}.


\begin{table*}
	\caption{Estimated costs and values of the features of the PMS-II.}
	\label{table_cost_value}
	\centering
	\input{table_cost_value}
\end{table*}

\begin{table*}
	\caption{Qualities and strengths of value-related dependencies among the features of the PMS-II.}
	\label{table_rho}
	\centering
	\input{table_rho}
\end{table*}

\begin{figure*}[!b]
	\begin{center}
		\subfigure[$\beta=0.0$]{%
			\label{fig_cs_result_00}
			\includegraphics[scale=0.48]{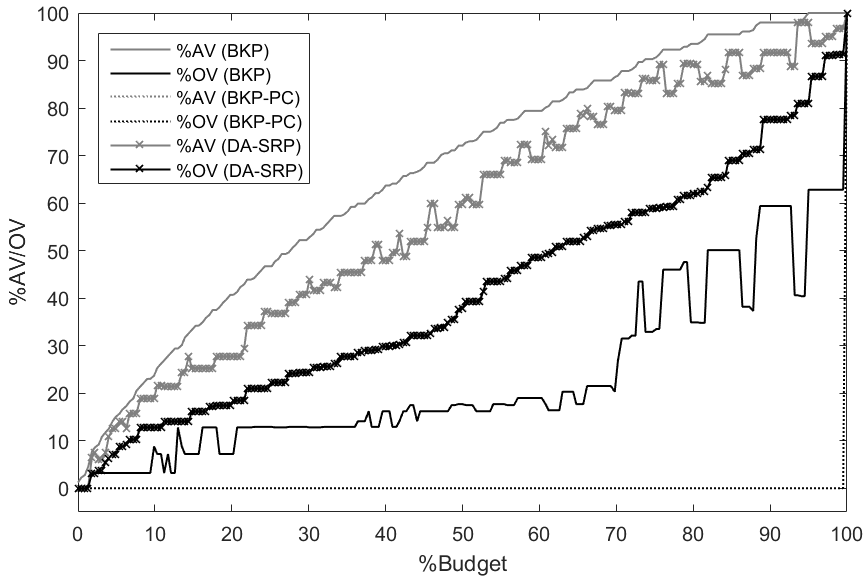}
		}
		\subfigure[$\beta=0.25$]{%
			\label{fig_cs_result_025}
			\includegraphics[scale=0.48]{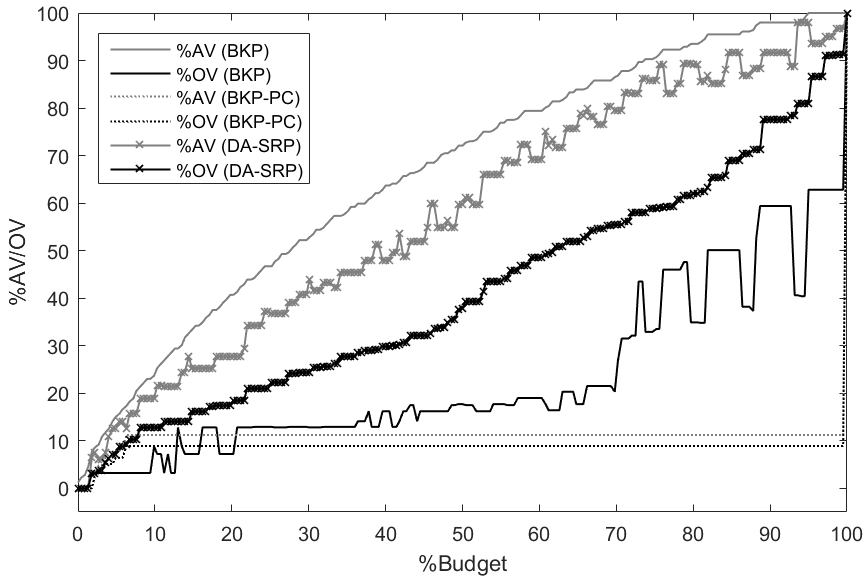}
		}
		\subfigure[$\beta=0.5$]{%
			\label{fig_cs_result_05}
			\includegraphics[scale=0.48]{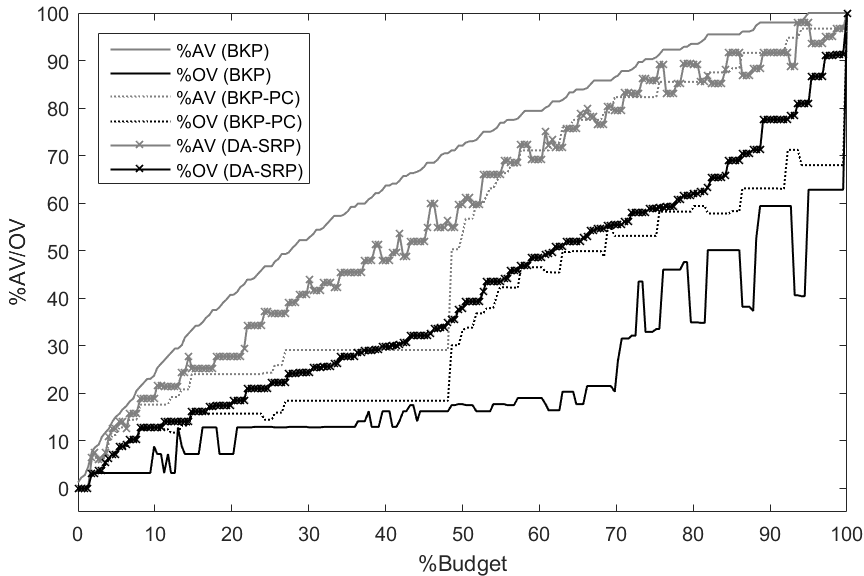}
		}
		\subfigure[$\beta=0.75$]{%
			\label{fig_cs_result_075}
			\includegraphics[scale=0.48]{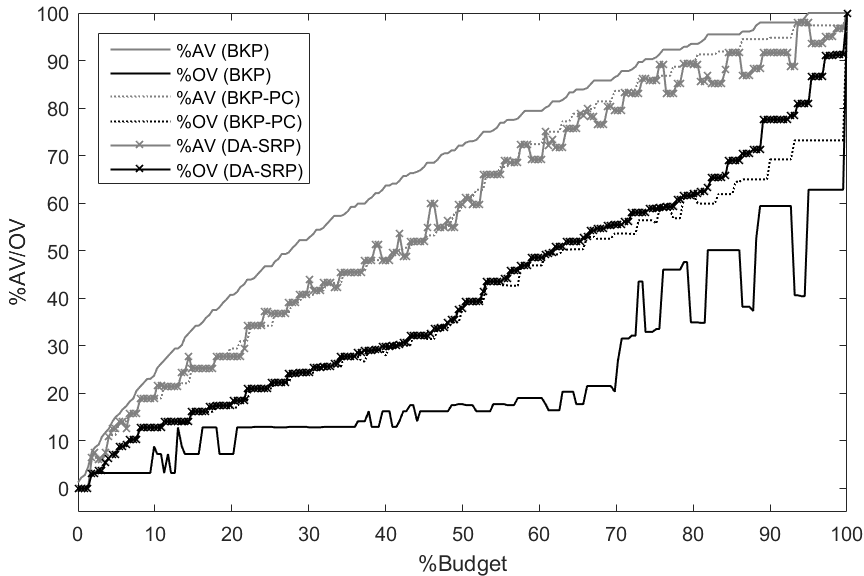}
		}
	\end{center}
	\caption{%
		Accumulated Value and Overall Value achieved for the features of the PMS-II using the BKP, BKP-PC, and DA-SRP models.
	}%
	\label{fig_cs_result}
\end{figure*}

Table~\ref{table_cost_value} lists the estimated costs and values of the PMS-II features. Also, Table~\ref{table_rho} lists the qualities ($+$/$-$/$\pm$) and strengths of value-related dependencies among those features. For a feature $f_i$ its corresponding row in Table \ref{table_rho} denotes the qualities and strengths of dependencies from $f_i$ to all other features of the PMS-II. 

Release planning was performed for the features of the PMS-II using the BKP, BKP-PC, and DA-SRP models to evaluate their performance in the presence of various budget constraints ($Budget \in \{1,...,222\}$). Figure \ref{fig_cs_result} summarizes the results of our experiments by comparing percentage of the \textit{accumulated value} ($\%AV=\frac{AV}{\sum_{i=1}^{27}v_i}$) and percentage of the overall value ($\%OV=\frac{OV}{\sum_{i=1}^{27}\phi_iv_i}$) achieved by each release planning model in the presence of various budget constraints. The horizontal axis shows percentages of the available budget ($\%Budget=\frac{Budget}{\sum_{i=1}^{27}c_i}, Budget=\{1,2,...,222\}$) and the vertical axis demonstrates percentages of accumulated value ($\%AV$) or percentages of overall value ($\%OV$) achieved by the release planning models. 

The results of our experiments (Figure \ref{fig_cs_result}) verified that the BKP model always maximized the accumulated value of the selected features (optimal subset of features) while the DA-SRP model maximized the overall value of those features. Also, as expected, overall values provided by the BKP-PC model for various budgets were equal to their corresponding accumulated values. The reason is that the BKP-PC model formulates value-related dependencies of all strengths as precedence constraints. As such, a feature $f_i$ with a positive (negative) dependency to a feature $f_j$ will not be selected unless $f_j$ is selected (ignored). Hence, the penalty of ignoring/selecting positive/negative dependencies of $f_i$ will always be zero.  


We also observed maximizing the accumulated value and the overall value of the features can be conflicting objectives as increasing one may reduce the other. This can be seen at many points in the graphs of the BKP model in Figure~\ref{fig_cs_result} where a budget increase has resulted in increasing the accumulated value while the overall value has decreased. For the DA-SRP model however, it can be observed in some pints of its corresponding graphs that budget increase has resulted in increasing the overall value while the accumulated value has decreased.

The BKP-PC model on the other hand, did not give any value (Figure~\ref{fig_cs_result_00}) even for higher budget percentages. This is for the reason that the BKP-PC model suffers from the selection deficiency problem (SDP)~\cite{mougouei2016factoring} as it can only formulate value-related dependencies as precedence constraints ignoring the strengths of those dependencies. As a result, a feature $f_i$ ($x_i=0$) can not be selected unless all of its positive (negative) dependencies (of any strength) are selected (ignored). This may result in ignoring a large number of features even in the presence of sufficient budget. 

\vspace{1em}

The adverse impact of the selection deficiency problem nevertheless, can be mitigated when only value-related dependencies whose strengths are beyond a certain $\beta \in [0,1]$ threshold ($\rho_{i,j} >\beta$) are formulated as precedence constraints. In this way, the BKP-PC model is formulated as a function of $\beta$ where greater values of $\beta$ ($\beta \rightarrow 1$) generally decrease the adverse impact of the \textit{SDP}.


Figures~\ref{fig_cs_result_025},~\ref{fig_cs_result_05},~\ref{fig_cs_result_075} compare performance of the BKP-PC model against that of the BKP and DA-SRP models when only value-related dependencies stronger than a certain $\beta$ threshold are formulated as precedence constraints by the BKP-PC model. As demonstrated in these figures, percentage of overall value ($\%OV$) achieved by the BKP-PC model was generally (not always) higher than the $\%OV$ achieved by the BKP model. The DA-SRP model however, outperformed the BKP-PC model by providing a greater or equal $\%OV$ for all budget values.

%

%% file: table_cost_value.tex
\huge\resizebox {1\textwidth }{!}{
\begin{tabular}{lllllllllllllllllllllllllllll}
\toprule[1.5pt]
\textbf{\cellcolor{black}\textcolor{white}{Feature}}&
\textbf{\cellcolor{black}\textcolor{white}{$f_{1}$}}&
\textbf{\cellcolor{black}\textcolor{white}{$f_{2}$}}&
\textbf{\cellcolor{black}\textcolor{white}{$f_{3}$}}&
\textbf{\cellcolor{black}\textcolor{white}{$f_{4}$}}&
\textbf{\cellcolor{black}\textcolor{white}{$f_{5}$}}&
\textbf{\cellcolor{black}\textcolor{white}{$f_{6}$}}&
\textbf{\cellcolor{black}\textcolor{white}{$f_{7}$}}&
\textbf{\cellcolor{black}\textcolor{white}{$f_{8}$}}&
\textbf{\cellcolor{black}\textcolor{white}{$f_{9}$}}&
\textbf{\cellcolor{black}\textcolor{white}{$f_{10}$}}&
\textbf{\cellcolor{black}\textcolor{white}{$f_{11}$}}&
\textbf{\cellcolor{black}\textcolor{white}{$f_{12}$}}&
\textbf{\cellcolor{black}\textcolor{white}{$f_{13}$}}&
\textbf{\cellcolor{black}\textcolor{white}{$f_{14}$}}&
\textbf{\cellcolor{black}\textcolor{white}{$f_{15}$}}&
\textbf{\cellcolor{black}\textcolor{white}{$f_{16}$}}&
\textbf{\cellcolor{black}\textcolor{white}{$f_{17}$}}&
\textbf{\cellcolor{black}\textcolor{white}{$f_{18}$}}&
\textbf{\cellcolor{black}\textcolor{white}{$f_{19}$}}&
\textbf{\cellcolor{black}\textcolor{white}{$f_{20}$}}&
\textbf{\cellcolor{black}\textcolor{white}{$f_{21}$}}&
\textbf{\cellcolor{black}\textcolor{white}{$f_{22}$}}&
\textbf{\cellcolor{black}\textcolor{white}{$f_{23}$}}&
\textbf{\cellcolor{black}\textcolor{white}{$f_{24}$}}&
\textbf{\cellcolor{black}\textcolor{white}{$f_{25}$}}&
\textbf{\cellcolor{black}\textcolor{white}{$f_{26}$}}&
\textbf{\cellcolor{black}\textcolor{white}{$f_{27}$}}&
\textbf{\cellcolor{black}\textcolor{white}{Sum}}
\\\midrule\textbf{\cellcolor{black}\textcolor{white}{Cost}}&
$5.0$&$20.0$&$0.0$&$10.0$&$1.0$&$20.0$&$6.0$&$5.0$&$16.0$&$10.0$&$4.0$&$3.0$&$5.0$&$7.0$&$15.0$&$13.0$&$14.0$&$3.0$&$10.0$&$7.0$&$12.0$&$15.0$&$8.0$&$2.0$&$10.0$&$0.0$&$1.0$&$222.0$
\\\midrule \textbf{\cellcolor{black}\textcolor{white}{Value}}&
$10.0$&$20.0$&$4.0$&$17.0$&$3.0$&$20.0$&$15.0$&$9.0$&$20.0$&$16.0$&$20.0$&$10.0$&$6.0$&$8.0$&$8.0$&$10.0$&$6.0$&$10.0$&$20.0$&$20.0$&$15.0$&$20.0$&$20.0$&$5.0$&$0.0$&$0.0$&$0.0$&$312.0$
\\\toprule[1.5pt]
\end{tabular}}

%% file: table_rho.tex
\huge\resizebox {1\textwidth }{!}{
\begin{tabular}{lllllllllllllllllllllllllllll}
\toprule[1.5pt]
\textbf{\cellcolor{black}\textcolor{white}{}}& \textbf{\cellcolor{black}\textcolor{white}{$f_{1}$}}
& \textbf{\cellcolor{black}\textcolor{white}{$f_{2}$}}
& \textbf{\cellcolor{black}\textcolor{white}{$f_{3}$}}
& \textbf{\cellcolor{black}\textcolor{white}{$f_{4}$}}
& \textbf{\cellcolor{black}\textcolor{white}{$f_{5}$}}
& \textbf{\cellcolor{black}\textcolor{white}{$f_{6}$}}
& \textbf{\cellcolor{black}\textcolor{white}{$f_{7}$}}
& \textbf{\cellcolor{black}\textcolor{white}{$f_{8}$}}
& \textbf{\cellcolor{black}\textcolor{white}{$f_{9}$}}
& \textbf{\cellcolor{black}\textcolor{white}{$f_{10}$}}
& \textbf{\cellcolor{black}\textcolor{white}{$f_{11}$}}
& \textbf{\cellcolor{black}\textcolor{white}{$f_{12}$}}
& \textbf{\cellcolor{black}\textcolor{white}{$f_{13}$}}
& \textbf{\cellcolor{black}\textcolor{white}{$f_{14}$}}
& \textbf{\cellcolor{black}\textcolor{white}{$f_{15}$}}
& \textbf{\cellcolor{black}\textcolor{white}{$f_{16}$}}
& \textbf{\cellcolor{black}\textcolor{white}{$f_{17}$}}
& \textbf{\cellcolor{black}\textcolor{white}{$f_{18}$}}
& \textbf{\cellcolor{black}\textcolor{white}{$f_{19}$}}
& \textbf{\cellcolor{black}\textcolor{white}{$f_{20}$}}
& \textbf{\cellcolor{black}\textcolor{white}{$f_{21}$}}
& \textbf{\cellcolor{black}\textcolor{white}{$f_{22}$}}
& \textbf{\cellcolor{black}\textcolor{white}{$f_{23}$}}
& \textbf{\cellcolor{black}\textcolor{white}{$f_{24}$}}
& \textbf{\cellcolor{black}\textcolor{white}{$f_{25}$}}
& \textbf{\cellcolor{black}\textcolor{white}{$f_{26}$}}
& \textbf{\cellcolor{black}\textcolor{white}{$f_{27}$}}
\\\midrule \addlinespace[0.4ex]
\textbf{\cellcolor{black}\textcolor{white}{$f_{1}$}}& $+1.00$& $+0.14$& $+0.09$& $+0.03$& $+0.14$& $+0.01$& $+0.01$& $+0.16$& $+0.03$& $+0.03$& $-0.02$& $+0.11$& $+0.03$& $+0.11$& $+0.15$& $+0.09$& $+0.01$& $-0.04$& $-0.04$& $-0.04$& $-0.02$& $+0.06$& $+0.14$& $+0.14$& $-0.14$& $-0.18$& $-0.01$\\
\textbf{\cellcolor{black}\textcolor{white}{$f_{2}$}}& \cellcolor{gray!30}$+1.00$& $+1.00$& $+0.20$& $+0.47$& $+1.00$& $+0.07$& $+0.02$& $+0.25$& $+0.47$& $+0.47$& $+0.25$& $+0.07$& $+0.47$& $+0.22$& $+0.24$& $+0.43$& $+0.22$& $+0.24$& $+0.12$& $+0.17$& $+0.18$& $+0.19$& $+1.00$& $+1.00$& $+0.18$& \cellcolor{gray!30}$+1.00$& $+0.15$\\
\textbf{\cellcolor{black}\textcolor{white}{$f_{3}$}}& \cellcolor{gray!30}$+1.00$& $+0.37$& $+1.00$& $+0.40$& $+0.37$& $-0.00$& $+0.15$& $+0.47$& $+0.40$& $+0.40$& $+0.33$& $-0.15$& $+0.40$& $+0.68$& $+0.50$& $+0.35$& $-0.07$& $+0.07$& $+0.01$& $+0.09$& $+0.48$& $+0.33$& $+0.37$& $+0.37$& $+0.00$& $-0.06$& $+0.02$\\
\textbf{\cellcolor{black}\textcolor{white}{$f_{4}$}}& $+0.05$& \cellcolor{gray!30}$+1.00$& $+0.32$& $+1.00$& $+0.72$& $+0.30$& $-0.02$& $+0.42$& $+1.00$& $+1.00$& $+0.39$& $+0.03$& $+1.00$& $+0.33$& $+0.39$& $+0.40$& $+0.12$& $+0.37$& $-0.00$& $+0.25$& $+0.27$& $+0.30$& $+0.72$& $+0.72$& $+0.27$& $-0.11$& $+0.24$\\
\textbf{\cellcolor{black}\textcolor{white}{$f_{5}$}}& \cellcolor{gray!30}$+1.00$& $+1.00$& $+0.20$& $+0.47$& $+1.00$& $+0.07$& $+0.02$& $+0.25$& $+0.47$& $+0.47$& $+0.25$& $+0.07$& $+0.47$& $+0.22$& $+0.24$& $+0.43$& $+0.22$& $+0.24$& $+0.12$& $+0.17$& $+0.18$& $+0.19$& $+1.00$& $+1.00$& $+0.18$& $+0.00$& $+0.15$\\
\textbf{\cellcolor{black}\textcolor{white}{$f_{6}$}}& $+0.02$& $+0.11$& $-0.00$& $+0.33$& $+0.11$& $+1.00$& $-0.05$& $+0.13$& $+0.33$& $+0.33$& $+0.14$& $+0.32$& $+0.33$& $+0.05$& $+0.09$& $-0.03$& $-0.19$& $+0.08$& $-0.37$& $+0.27$& $-0.25$& $-0.03$& $+0.11$& $+0.11$& $+0.04$& \cellcolor{gray!30}$+1.00$& $+0.30$\\
\textbf{\cellcolor{black}\textcolor{white}{$f_{7}$}}& $+0.01$& $+0.04$& $+0.11$& $-0.02$& $+0.04$& $-0.04$& $+1.00$& $-0.11$& $-0.02$& $-0.02$& $+0.21$& $-0.14$& $-0.02$& $+0.12$& $-0.14$& $-0.04$& $-0.10$& $-0.16$& $-0.09$& $+0.17$& $-0.03$& $-0.23$& $+0.04$& $+0.04$& $-0.04$& $+0.02$& $+0.16$\\
\textbf{\cellcolor{black}\textcolor{white}{$f_{8}$}}& \cellcolor{gray!30}$+1.00$& $+0.46$& $+0.47$& $+0.51$& $+0.46$& $+0.14$& $-0.15$& $+1.00$& $+0.51$& $+0.51$& $+0.24$& $-0.07$& $+0.51$& $+0.54$& $+0.70$& $+0.25$& $+0.01$& $+0.21$& $-0.11$& $-0.09$& $+0.36$& $+0.43$& $+0.46$& $+0.46$& \cellcolor{gray!30}$+1.00$& $-0.24$& $-0.00$\\
\textbf{\cellcolor{black}\textcolor{white}{$f_{9}$}}& $+0.05$& $+0.72$& $+0.32$& $+1.00$& $+0.72$& \cellcolor{gray!30}$+1.00$& $-0.02$& $+0.42$& $+1.00$& $+1.00$& $+0.39$& $+0.03$& $+1.00$& $+0.33$& $+0.39$& $+0.40$& $+0.12$& $+0.37$& $-0.00$& $+0.25$& $+0.27$& $+0.30$& $+0.72$& $+0.72$& $+0.27$& $-0.11$& $+0.24$\\
\textbf{\cellcolor{black}\textcolor{white}{$f_{10}$}}& $+0.05$& $+0.72$& $+0.32$& $+1.00$& $+0.72$& $+0.30$& $-0.02$& $+0.42$& $+1.00$& $+1.00$& $+0.39$& $+0.03$& $+1.00$& $+0.33$& $+0.39$& $+0.40$& $+0.12$& $+0.37$& $-0.00$& $+0.25$& $+0.27$& $+0.30$& $+0.72$& $+0.72$& $+0.27$& $-0.11$& $+0.24$\\
\textbf{\cellcolor{black}\textcolor{white}{$f_{11}$}}& $-0.04$& $+0.49$& $+0.34$& $+0.49$& $+0.49$& $+0.16$& $+0.29$& $+0.25$& $+0.49$& $+0.49$& $+1.00$& $-0.04$& $+0.49$& $+0.42$& $+0.20$& $+0.26$& $+0.10$& $+0.28$& $+0.00$& $+0.42$& $+0.36$& $-0.02$& $+0.49$& $+0.49$& $+0.24$& $-0.06$& $+0.24$\\
\textbf{\cellcolor{black}\textcolor{white}{$f_{12}$}}& \cellcolor{gray!30}$+1.00$& $+0.08$& $-0.10$& $+0.03$& $+0.08$& $+0.22$& $-0.11$& $-0.04$& $+0.03$& $+0.03$& $-0.02$& $+1.00$& $+0.03$& $-0.10$& $-0.06$& $+0.02$& $-0.02$& $-0.06$& $-0.04$& $+0.17$& $-0.26$& $-0.11$& $+0.08$& $+0.08$& $+0.01$& $-0.20$& $+0.17$\\
\textbf{\cellcolor{black}\textcolor{white}{$f_{13}$}}& $+0.05$& $+0.72$& $+0.32$& $+1.00$& $+0.72$& \cellcolor{gray!30}$+1.00$& $-0.02$& $+0.42$& $+1.00$& $+1.00$& $+0.39$& $+0.03$& $+1.00$& $+0.33$& $+0.39$& $+0.40$& $+0.12$& $+0.37$& $-0.00$& $+0.25$& $+0.27$& $+0.30$& $+0.72$& $+0.72$& $+0.27$& $-0.11$& $+0.24$\\
\textbf{\cellcolor{black}\textcolor{white}{$f_{14}$}}& \cellcolor{gray!30}$+1.00$& $+0.43$& $+0.71$& $+0.43$& $+0.43$& $+0.06$& $+0.17$& $+0.57$& $+0.43$& $+0.43$& $+0.42$& $-0.16$& $+0.43$& $+1.00$& $+0.59$& $+0.37$& $-0.00$& $+0.14$& $-0.11$& $+0.10$& $+0.46$& $+0.27$& $+0.43$& $+0.43$& $-0.06$& $-0.18$& $-0.03$\\
\textbf{\cellcolor{black}\textcolor{white}{$f_{15}$}}& $+0.32$& \cellcolor{gray!30}$+1.00$& $+0.52$& $+0.50$& $+0.47$& $+0.10$& $-0.20$& $+0.73$& $+0.50$& $+0.50$& $+0.20$& $-0.09$& $+0.50$& $+0.58$& $+1.00$& $+0.48$& $+0.10$& $+0.30$& $-0.01$& $-0.04$& \cellcolor{gray!30}$+1.00$& $+0.51$& $+0.47$& $+0.47$& $+0.02$& $-0.17$& $-0.13$\\
\textbf{\cellcolor{black}\textcolor{white}{$f_{16}$}}& $+0.14$& \cellcolor{gray!30}$+1.00$& $+0.25$& $+0.36$& $+0.59$& $-0.02$& $-0.04$& $+0.18$& $+0.36$& $+0.36$& $+0.18$& $+0.02$& $+0.36$& $+0.26$& $+0.33$& $+1.00$& $+0.31$& $+0.30$& $+0.25$& $+0.20$& $+0.25$& $+0.24$& $+0.59$& $+0.59$& $+0.23$& $+0.08$& $-0.01$\\
\textbf{\cellcolor{black}\textcolor{white}{$f_{17}$}}& \cellcolor{gray!30}$+1.00$& $+0.41$& $-0.07$& $+0.15$& $+0.41$& $-0.22$& $-0.13$& $+0.01$& $+0.15$& $+0.15$& $+0.10$& $-0.03$& $+0.15$& $-0.00$& $+0.10$& $+0.44$& $+1.00$& $+0.52$& $+0.49$& $+0.12$& $+0.33$& $+0.24$& $+0.41$& $+0.41$& $+0.42$& $+0.32$& $-0.01$\\
\textbf{\cellcolor{black}\textcolor{white}{$f_{18}$}}& \cellcolor{gray!30}$+1.00$& $+0.48$& $+0.08$& $+0.47$& $+0.48$& $+0.09$& $-0.23$& $+0.22$& $+0.47$& $+0.47$& $+0.28$& $-0.09$& $+0.47$& $+0.14$& $+0.30$& $+0.44$& \cellcolor{gray!30}$+1.00$& $+1.00$& $+0.29$& $+0.22$& $+0.46$& $+0.39$& $+0.48$& $+0.48$& $+0.57$& $+0.16$& $+0.07$\\
\textbf{\cellcolor{black}\textcolor{white}{$f_{19}$}}& $-0.08$& \cellcolor{gray!30}$+1.00$& $+0.01$& $-0.00$& $+0.21$& \cellcolor{gray!30}$+1.00$& $-0.11$& $-0.11$& \cellcolor{gray!30}$+1.00$& $-0.00$& $+0.00$& $-0.06$& $-0.00$& $-0.10$& $-0.01$& $+0.33$& $+0.47$& $+0.26$& $+1.00$& $+0.03$& $+0.31$& $+0.17$& $+0.21$& $+0.21$& $+0.35$& $+0.49$& $-0.11$\\
\textbf{\cellcolor{black}\textcolor{white}{$f_{20}$}}& $-0.07$& \cellcolor{gray!30}$+1.00$& $+0.08$& $+0.26$& $+0.26$& \cellcolor{gray!30}$+1.00$& $+0.19$& $-0.07$& \cellcolor{gray!30}$+1.00$& $+0.26$& $+0.33$& $+0.22$& $+0.26$& $+0.08$& $-0.03$& $+0.23$& $+0.10$& $+0.17$& $+0.03$& $+1.00$& $+0.01$& $-0.10$& $+0.26$& $+0.26$& $+0.36$& $-0.05$& $+0.42$\\
\textbf{\cellcolor{black}\textcolor{white}{$f_{21}$}}& $-0.03$& \cellcolor{gray!30}$+1.00$& $+0.33$& $+0.22$& $+0.23$& $-0.19$& $-0.03$& $+0.25$& $+0.22$& $+0.22$& $+0.24$& $-0.28$& $+0.22$& $+0.30$& $+0.28$& $+0.24$& $+0.23$& $+0.30$& $+0.22$& $+0.01$& $+1.00$& $+0.33$& $+0.23$& $+0.23$& $+0.23$& $+0.22$& $-0.09$\\
\textbf{\cellcolor{black}\textcolor{white}{$f_{22}$}}& \cellcolor{gray!30}$+1.00$& $+0.34$& $+0.32$& $+0.35$& $+0.34$& $-0.03$& $-0.29$& $+0.42$& $+0.35$& $+0.35$& $-0.01$& $-0.16$& $+0.35$& $+0.25$& $+0.48$& $+0.33$& $+0.23$& $+0.36$& $+0.17$& $-0.12$& $+0.47$& $+1.00$& $+0.34$& $+0.34$& $+0.24$& $+0.11$& $-0.01$\\
\textbf{\cellcolor{black}\textcolor{white}{$f_{23}$}}& $+0.16$& \cellcolor{gray!30}$+1.00$& $+0.20$& $+0.47$& $+1.00$& $+0.07$& $+0.02$& $+0.25$& $+0.47$& $+0.47$& $+0.25$& $+0.07$& $+0.47$& $+0.22$& $+0.24$& $+0.43$& $+0.22$& $+0.24$& $+0.12$& $+0.17$& $+0.18$& $+0.19$& $+1.00$& $+1.00$& $+0.18$& $+0.00$& $+0.15$\\
\textbf{\cellcolor{black}\textcolor{white}{$f_{24}$}}& \cellcolor{gray!30}$+1.00$& $+1.00$& $+0.20$& $+0.47$& $+1.00$& $+0.07$& $+0.02$& $+0.25$& $+0.47$& $+0.47$& $+0.25$& $+0.07$& $+0.47$& $+0.22$& $+0.24$& $+0.43$& $+0.22$& $+0.24$& $+0.12$& $+0.17$& $+0.18$& $+0.19$& $+1.00$& $+1.00$& $+0.18$& $+0.00$& $+0.15$\\
\textbf{\cellcolor{black}\textcolor{white}{$f_{25}$}}& \cellcolor{gray!30}$+1.00$& $+0.29$& $+0.00$& $+0.28$& $+0.29$& $+0.04$& $-0.04$& $-0.04$& $+0.28$& $+0.28$& $+0.20$& $+0.01$& $+0.28$& $-0.05$& $+0.02$& $+0.27$& $+0.36$& $+0.47$& $+0.32$& $+0.38$& $+0.29$& $+0.21$& $+0.29$& $+0.29$& $+1.00$& $+0.30$& $+0.27$\\
\textbf{\cellcolor{black}\textcolor{white}{$f_{26}$}}& \cellcolor{gray!30}$+1.00$& $+0.01$& $-0.06$& $-0.13$& $+0.01$& $-0.53$& $+0.02$& $-0.24$& $-0.13$& $-0.13$& $-0.06$& $-0.30$& $-0.13$& $-0.17$& $-0.16$& $+0.11$& $+0.31$& $+0.15$& $+0.50$& $-0.06$& $+0.31$& $+0.11$& $+0.01$& $+0.01$& $+0.34$& $+1.00$& \cellcolor{gray!30}$+1.00$\\
\textbf{\cellcolor{black}\textcolor{white}{$f_{27}$}}& \cellcolor{gray!30}$+1.00$& $+0.25$& $+0.01$& $+0.27$& $+0.25$& $+0.30$& $+0.19$& $-0.00$& $+0.27$& $+0.27$& $+0.20$& $+0.24$& $+0.27$& $-0.02$& $-0.11$& $-0.01$& $-0.01$& $+0.06$& $-0.10$& $+0.46$& $-0.12$& $-0.01$& $+0.25$& $+0.25$& $+0.28$& $-0.10$& $+1.00$\\
\addlinespace[0.4ex]\toprule[1.5pt]
\end{tabular}}

%% file: conclusion.tex
\section{Conclusions and Future Work}
\label{sec_conclusion}

In this paper we proposed finding value-related dependencies among software features through mining user preferences and then considering those value-related dependencies in software release planning. 

In this regard, a semi-automated approach was presented for identification of value-related dependencies and their characteristics (quality and strength). We then, used algebraic structure of fuzzy graphs for modeling value-related dependencies. Moreover, we demonstrated an application of a resampling technique for generating user preferences when the number of samples is not sufficient. For this purpose, we used a Latent Multivariate Gaussian model for generating large samples of user preferences from the initial sample directly gathered from users.

Finally, we formulated an integer programming model for dependency-aware release planning (DA-SRP) that finds an optimal subset of software features with the highest overall value where overall value captures the impacts of value-related dependencies among features. The validity and practicality of the work were demonstrated through studying a real world software project. 

The present work can be extended to explore various measures of casual strength and compare their efficiency in capturing value-related dependencies among software features. The work can also be continued by applying the presented approach to publicly available software repositories for gathering user preferences and perform dependency-aware release planning. Feedback from stakeholders (developers) then, can be studied and considered by release planning models.   
\vspace{1em}